\documentclass[aps,prb,preprint,superscriptaddress,notitlepage]{revtex4-2}
\usepackage[version=3]{mhchem}
\usepackage{graphicx} 
\usepackage{tabularx}
\usepackage{float}
\usepackage[english]{babel}
\addto{\captionsenglish}{%
  
}

\usepackage{array}
\usepackage{droidsans}
\usepackage{charter}
\usepackage[T1]{fontenc}
\usepackage[usenames,dvipsnames]{xcolor}
\usepackage{hyperref}
\usepackage{epstopdf}%

\definecolor{cream}{RGB}{222,217,201}

\begin{document}

\title{Electronic and structural properties of V$_2$O$_5$ layered polymorphs}

\author{Sakthi Kasthurirengan}
\email{skasthuri@ku.edu}
\affiliation{Department of Physics and Astronomy, University of Kansas, Lawrence, KS 66045, USA}

\author{Hartwin Peelaers}
\email{peelaers@ku.edu}
\affiliation{Department of Physics and Astronomy, University of Kansas, Lawrence, KS 66045, USA}

\begin{abstract}
V$_2$O$_5$ is a promising battery electrode material that can intercalate not only Li, but also more abundant alkaline metals such as Na and K, and even multivalent ions such as Al, Ca, Cu, Mg, and Zn. V$_2$O$_5$ exhibits several different polymorphs, and phase transitions between the polymorphs can occur depending on intercalant or external conditions. At least 8 different layered polymorphs have been observed. However, detailed information about the energetics and structural properties of each polymorph is still lacking. 
To obtain a reliable computational reference, we use hybrid density functional theory calculations to investigate the properties of layered V$_2$O$_5$ polymorphs. We benchmarked several methods to include van der Waals interactions in combination with hybrid functionals, and found that the Grimme D3 method is most accurate. We obtain detailed information on the electronic properties and structures of the various unintercalated polymorphs and show that the main electronic effect of intercalants is a filling of the lowest conduction bands, as the intercalant contributions are well above the conduction-band minimum. Despite the structural differences between the unintercalated polymorphs, we find that they have very similar band gaps and band structures, with the exception of the high temperature and pressure phase $\beta$.
\end{abstract}

\maketitle
\newpage

Lithium-ion batteries have become a ubiquitous part of our increasingly technologically-dependent lives. However, the growing demand for such batteries puts a large strain on available resources for both Li and the cathode materials. The high cost, coupled with safety concerns, also prevents wide-spread grid-scale usage. 

Vanadium pentoxide, V$_2$O$_5$, is a viable alternative candidate cathode material \cite{Whittingham2004}, as 
vanadium is abundant in the earth's crust \cite{Simandl2009,Raja2007}. It can reliably intercalate different intercalants, including Li~\cite{Delmas1994,Christensen2019}, Na~\cite{Tang2004,Etman2018,Emery2018}, K~\cite{Tang2004,Clites2018}, Mg~\cite{Tang2004,Attias2018,Attias2019,Ming2018,Le1998, Gershinsky2013, Verrelli2018}, Zn~\cite{Yan2018,Senguttuvan2016,Zhang2018a,Le1998,Chen2019,Bhatia2022, Hao2020, Guo2018}, and Al~\cite{Le1998}.
Supercapacitors \cite{Marley2015a, Chen2011}, gaseous sensor materials \cite{Modafferi2012,Suresh2014}, and light-rechargeable Li-ion batteries \cite{Boruah2021} have been demonstrated. 
There are several
chemical and physical synthesis routes to obtain V$_2$O$_5$, including RF and DC sputtering\cite{Ottaviano2004,Navone2005}, pulsed laser deposition\cite{Julien1999,Ramana2005}, chemical vapor
deposition\cite{Drosos2018}, atomic layer deposition\cite{Chen2012}, electrodeposition\cite{Lee2009}, thermal evaporation\cite{Santos2013,Cheng2006}, thermal oxidation\cite{Swiatowska-Mrowiecka2007}, and the sol–gel method \cite{Gotic2003}, making V$_2$O$_5$ an appealing option for future batteries. The various synthesis routes and the inclusion of different intercalants leads to a wide variety of structural polymorphs which complicates the understanding of V$_2$O$_5$ fundamental properties.

\begin{figure}[htb]
	\centering
	\includegraphics[width=0.7\columnwidth]{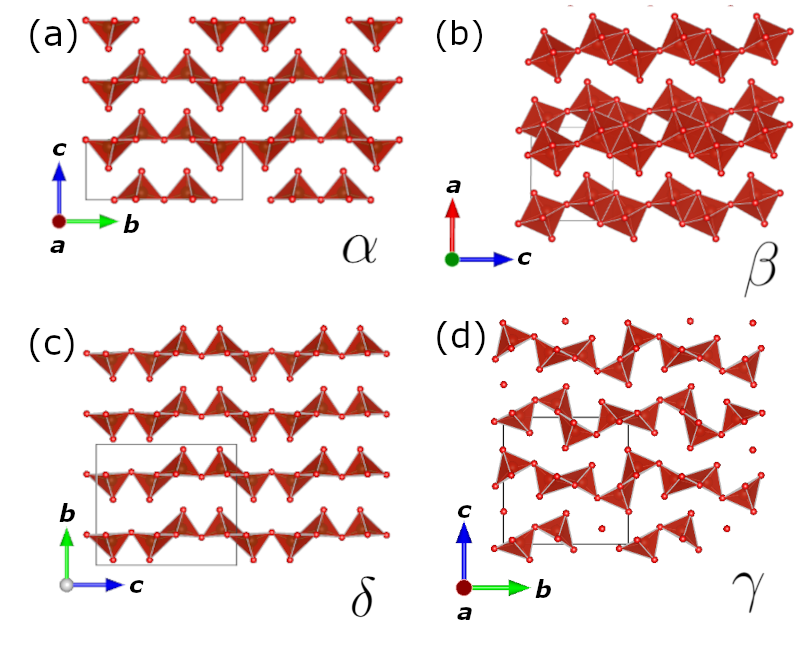}
	\caption{The single-layer polymorphs investigated in this study: (a) $\alpha$-V$_2$O$_5$, (b) $\beta$-V$_2$O$_5$, (c) $\delta$-V$_2$O$_5$, and (d) $\gamma$-V$_2$O$_5$. Polyhedra are shown around the V atoms (large spheres) to indicate the bonding environment. Smaller spheres are O atoms.}
	\label{singles}
\end{figure}

In ambient conditions V$_2$O$_5$ stabilizes in an orthorhombic crystal structure, $\alpha$-V$_2$O$_5$, with space group \emph{Pmmn}. $\alpha$-V$_2$O$_5$ is composed of weakly bonded layers of V$_2$O$_5$ square pyramids as shown in Fig.~\ref{singles}(a). For high temperatures and pressures, the layered $\beta$ polymorph (space group \emph{P2$_1$/m}) is found~\cite{Balog2007, Filonenko2001} [Fig.~\ref{singles}(b)]. 
Upon Li intercalation (Li$_x$V$_2$O$_5$) several phase transitions take place
\cite{Whittingham2004,Galy1992,Baddour-Hadjean2006}. The $\alpha-$ [Fig.~\ref{singles}(a)] and $\epsilon-$ structures maintain an orthorhombic (\emph{Pmmn}) layered structure for concentration of $0 < x \leq 0.7$, with these phase transitions being reversible under delithiation \cite{Smirnov2018, Baddour-Hadjean2006}. The $\alpha$ polymorph is the ground state and is stable for $0 < x < 0.1$. The $\epsilon$ polymorph occurs for $0.35 < x < 0.7$. We were unable to stabilize this polymorph without the presence of Li, so it is not shown in Fig.~\ref{singles}. 
The $\delta$ polymorph [Fig.~\ref{singles}(c)] is observed for the range $0.7 < x < 1$ and has space group $Cmcm$ \cite{Baddour-Hadjean2006}. This polymorph is similar to the $\alpha$ phase except that each alternate layer is shifted along the $b$ axis or [010] plane by a distance of $b/2$  \cite{Galy1992}. 
It has also been observed when intercalating with Mg \cite{Bouloux1976, Onoda1998}.

For $1<x<2$,V$_2$O$_5$ occurs in the $\gamma$ polymorph (\emph{Pnma} space group) [Fig.~\ref{singles}(d)], which has been shown to be an irreversible transformation under delithiation \cite{Cocciantelli1995}. Initial studies of this polymorph showed high energy densities \cite{Baddour-Hadjean2019,Delmas1994}.

For $x > 2$, V$_2$O$_5$ occurs in the $\omega$ polymorph which is no longer layered as it has a rock-salt structure \cite{Whittingham2004}. This structure was initially thought to be an irreversible transformation \cite{Delmas1994}, but delithiation of this polymorph leads to a disordered structure called $\beta$-Li$_{0.3}$ V$_2$O$_5$~\cite{Christensen2019}. 

\begin{figure}[htb]
	\centering
	\includegraphics[width=0.7\columnwidth]{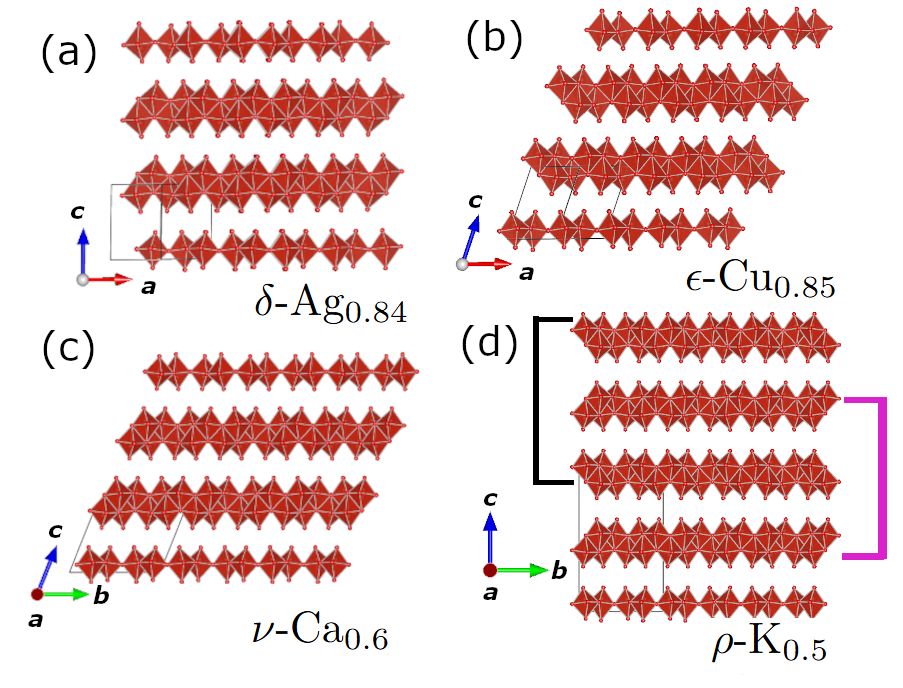}
	\caption{The double-layer polymorphs investigated in this study: (a) $\delta$-Ag$_{0.84}$-V$_2$O$_5$, (b) $\epsilon$-Cu$_{0.85}$-V$_2$O$_5$, (c) $\nu$-Ca$_{0.6}$-V$_2$O$_5$, and (d) $\rho$-K$_{0.5}$-V$_2$O$_5$. The black and purple indicators in (d) show that each alternating layer is identical and consists of D4 for the black layers and D4M for the purple layers.}
	\label{doubles}
\end{figure}

When V$_2$O$_5$ is intercalated with larger ions (compared to Li), so-called double-layer structures composed of distorted octahedra can be observed, as discussed in detail in Ref.~\onlinecite{Galy1992}. 
Fig.~\ref{doubles} shows 3 different double-layer structures, which differ in how the octahedra are stacked in each layer. The shift in the [100] direction of each of the layers is characterized in units of $Oc$=3.8\r{A} (the diagonal length across an octahedra)\cite{Galy1992}. 
The structure with a shift of 0 $Oc$ is called the $\delta$ polymorph or D4 (following Galy's notation\cite{Galy1992}) [see Fig.~\ref{doubles}(a)], which is observed when intercalating with Ag \cite{Garcia-Alvarado1992, Monchoux2011}. When one layer of the D4 structure is shifted by $+1$ $Oc$, the $\epsilon$ polymorph [Fig.~\ref{doubles}(b)] is obtained, observed with Cu intercalation\cite{Galy1970, Monchoux2011}. The D4 structure with a -0.5 $Oc$ shift is the $\nu$ polymorph [Fig.~\ref{doubles}(c)], which occurs under Ca intercalation~\cite{Kutoglu1983}. And finally, stacking alternate layers of D4 and D4M (the mirror image of the D4 structure) yields the $\rho$ polymorph [Fig.~\ref{doubles}(d)], which is observed with K intercalation~\cite{Savariault1992}. The $\delta$ and $\epsilon$ labels for these double-layer polymorphs are not the same as, or even related to, the structures of the $\delta$ and $\epsilon$ single-layer polymorphs. 
To avoid confusion caused by this historical naming convention we will suffix the double-layer polymorphs by their first observed intercalants (as reported in Ref.~\onlinecite{Galy1992} and references therein).

While most of these polymorphs have been reported in intercalated forms, unintercalated polymorphs can be synthesized experimentally by topotactic removal of ions using strong oxidizing agents. For example, single layer polymorphs $\alpha$ and $\gamma$ have been reported in a fully delithiated state \cite{Baddour-Hadjean2009}. $\beta$, which can accommodate sodium reversibly \cite{Cordoba2023}, can be produced unintercalated at high pressure-high temperature conditions \cite{Balog2007, Baddour-Hadjean2012}. The bilayered $\epsilon$-Cu phase formed with Cu in concentrations between 0.8-1, can be leached of all copper with an oxidizing treatment leaving the bilayered structure intact \cite{Baddour-Hadjean2018}.

Here, we explore the structural and energetic properties of the layered V$_2$O$_5$ polymorphs using first-principles calculations with hybrid functionals. We benchmark various methods of including van der Waals interactions in combination with hybrid functionals, finding that the Grimme D3 method performs best for this class of materials. We compare calculated structural parameters with experimentally reported values and analyze the (meta)stability of the polymorphs. We show that the electronic band structure of all layered polymorphs, despite the large structural differences, especially between single- and double-layered polymorphs, is remarkably similar. The magnitude of the band gap and the band characters of the valence and conduction bands are similar, and all exhibit a split-off conduction band (with the exception of the high temperature and pressure $\beta$ phase). Finally, we briefly discuss the electronic effects of intercalants. We show that their energy levels are located high in the conduction bands, so that their main electronic contribution is the addition of electrons to the split-off bands. This in turn leads to a lowering of the filled, or half-filled, conduction bands. 
Combined, these results form a systematic computational reference of layered V$_2$O$_5$ polymorphs.

\section{Computational Methods}

We use density functional theory (DFT) with projector augmented wave (PAW) potentials~\cite{Blochl1994} as implemented in the Vienna Ab-initio Simulation Package (VASP) \cite{Kresse1993,Kresse1996}. To obtain accurate structural and electronic properties we use the HSE06 hybrid functional~\cite{Heyd2003,Heyd2006}. A plane wave cutoff of 500 eV was used. All structures were relaxed so that the forces were smaller than 10 meV/\AA~and the stresses smaller than 0.6 meV/\AA$^{3}$.
We use a 6$\times$4$\times$2 {\bf k}-point grid for the $\alpha$ polymorph, and equivalent {\bf k}-point densities for the other polymorphs. 
The high-symmetry paths used for band structure calculations were obtained using the \textsc{AFLOW} package \cite{Curtarolo2012}. 
The site-projected band structures were plotted using the \textsc{pymatgen} python code \cite{Ong2013}. 
Structures are visualized using the \textsc{VESTA} code~\cite{Momma2011}. 

\subsection{van der Waals interactions}

\begin{figure}[tb]
	\centering
	\includegraphics[width=0.7\columnwidth]{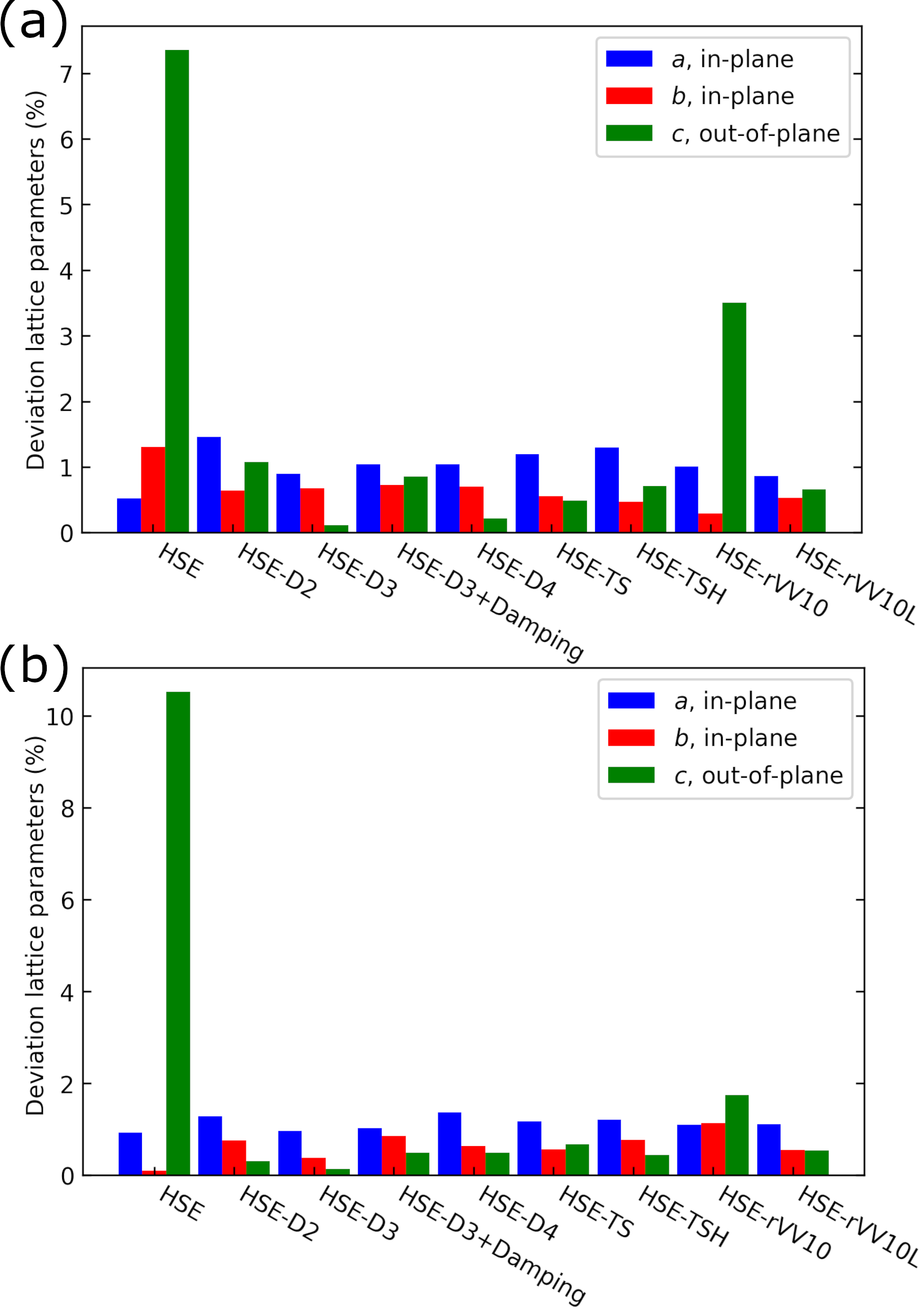}
	\caption{Deviation (in percent) of the calculated lattice constants as compared to the experimental lattice constants~\cite{Enjalbert1986, Balog2007}, for different computational methods~\cite{Grimme2006,Grimme2010,Grimme2011,Tkatchenko2009,Bucko2014} for the (a) $\alpha$ and (b) $\beta$ V$_2$O$_5$ polymorphs.}
	\label{vdwplot}
\end{figure}

Given the layered nature of the considered polymorphs, and the lack of van der Waals (vdW) interactions in standard DFT functionals, including the hybrid functional HSE06 used here, we benchmarked several methods to include vdW interactions. We restricted ourselves to methods implemented in the \textsc{VASP} code and only those that are compatible with HSE06~\cite{Grimme2006,Grimme2010,Grimme2011,Tkatchenko2009,Bucko2014}. In practice, that restricts us to methods that are based on an inclusion of a multipole expansion term to capture the vdW interactions~\cite{Bucko2013,Peelaers2014}. We used two different polymorphs ($\alpha$ and $\beta$) as test cases. The results are shown in Fig.~\ref{vdwplot}, where we compare the lattice constants to experimental data~\cite{Enjalbert1986, Balog2007}. The same data in tabular format can be found in the Supplementary Information(SI) as Table S1.
Not including any vdW interaction leads to a large error (7-10\%) in the out-of-plane lattice direction, clearly indicating the need for including the vdW interactions. With vdW interactions included, the out-of-plane lattice constants are described more accurately, independent of the used approach~\cite{Grimme2006,Grimme2010,Grimme2011,Caldeweyher2017,Tkatchenko2009,Bucko2014}.
The in-plane lattice constants do not depend strongly on the specific approach to include vdW interactions. Overall, the Grimme D3~\cite{Grimme2011} was the most accurate, hence we used this functional for all other calculations. We also tested the $B$ polymorph, which is not layered (see Table S2 in SI), to confirm that the inclusion of vdW interactions does not lead to errors in non-layered polymorphs, such as overbinding.

\section{Structural properties}

We fully relaxed 8 layered unintercalated polymorphs (4 single-layer and 4 double-layer) and listed the lattice constants in Table~\ref{lattConstTable}. We compare with available experimental results.~\cite{Enjalbert1986,Balog2007,Bouloux1976,Cocciantelli1995,Baddour-Hadjean2019,Rozier2009,Smirnov2018,Galy1992} 
Note that even for experimentally intercalated structures, we still obtain excellent agreement between our predictions and the observed lattice constants, which might indicate that the main role of the intercalants is the stabilization of the structure, while donating electrons to the conduction band. We will further validate this hypothesis in Section~\ref{sec:intercalated}. 

\begin{figure}[htb]
	\centering
	\includegraphics[width=0.7\columnwidth]{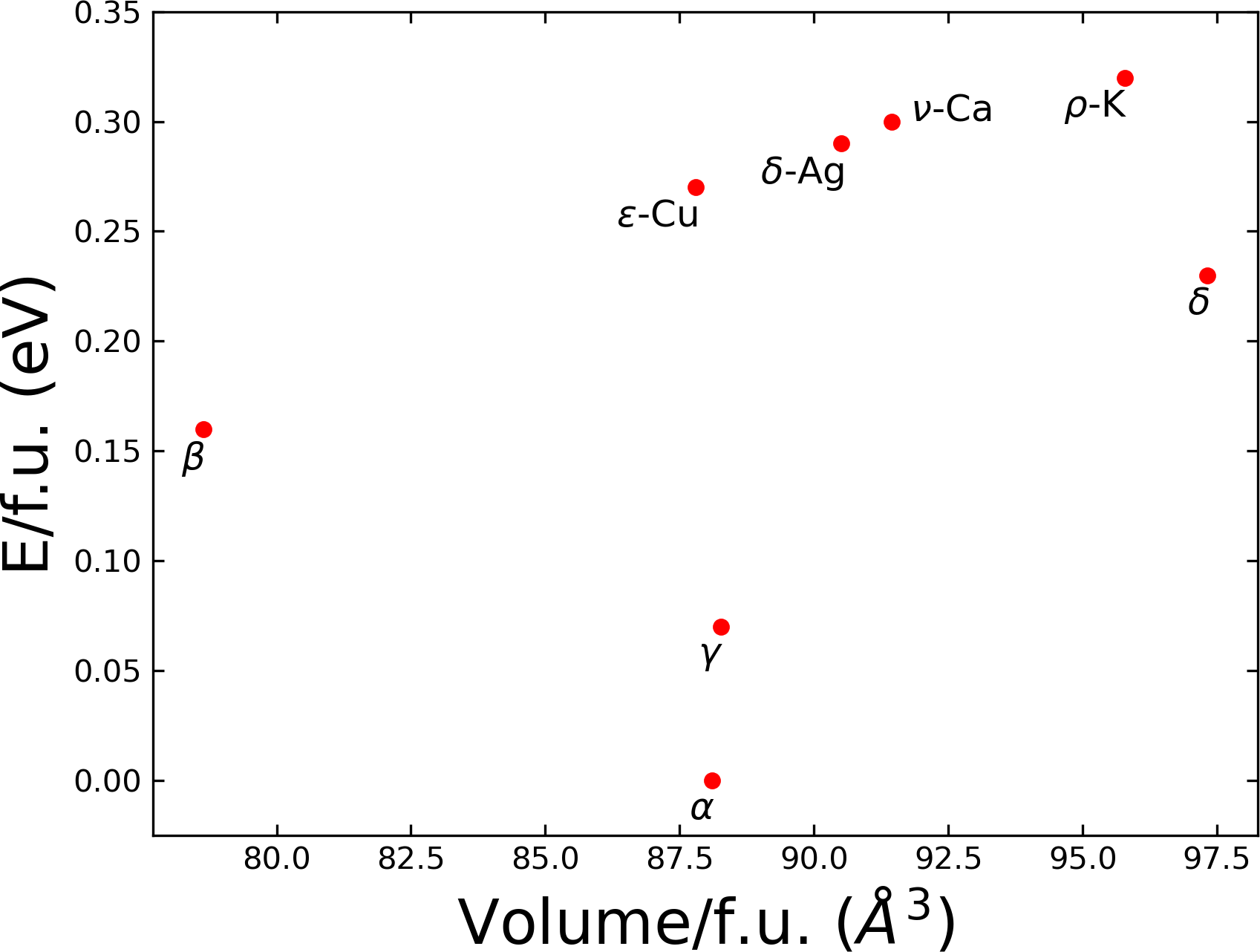}
	\caption{The energy per formula unit (f.u.) as function of volume per formula unit for various layered V$_2$O$_5$ polymorphs. The energy per formula unit of the $\alpha$ polymorph is set to 0.}
	\label{EvVol}
\end{figure}

Energetically, we find that the $\alpha$ polymorph is indeed the ground state structure. The $\gamma$ polymorph is 0.07 eV per formula unit (f.u.) higher in energy with a similar volume, while the other single-layer polymorphs are much higher in energy, 0.16 eV/f.u. for $\beta$ and 0.22 eV/f.u. for $\delta$, with large volume differences. The double-layered polymorphs are significantly higher in energy [at least 0.27 eV/f.u. (for $\epsilon$-Cu$_{0.85}$)], with $\epsilon$-Cu$_{0.85}$ having a similar volume to the single-layer $\alpha$ polymorph, and the other double-layered polymorphs having increased volume. These energetics and volumes are reported in Figure \ref{EvVol}.

\begin{table}[htb]
\caption{Calculated deintercalated lattice constants for the considered single-layer polymorphs ($\alpha$, $\beta$, $\delta$, and $\gamma$) and double-layer polymorphs ($\delta$-Ag$_{0.84}$, $\epsilon$-Cu$_{0.85}$, $\nu$-Ca$_{0.6}$, and $\rho$-K$_{0.5}$) as compared to experimental values. Layer direction lattice constant is labelled with a star (*).}
\begin{tabular}{cccccc}
  & a (\AA) & b (\AA) & c (\AA) & $\beta$ & Space \\
  &         &         &         & (deg.)  & Group \\
\hline
$\alpha$                  & 11.43  & 3.53 & 4.36*      & -           & $Pmmn$                          \\
Exp. $\alpha$\cite{Enjalbert1986}                   &11.51 & 3.56 & 4.37*      & -           & $Pmmn$                          \\
\hline
$\beta$               & 7.12* & 3.53    & 6.26         & 90.39   & $P2_1/m$          \\
Exp. $\beta$\cite{Balog2007}            & 7.11*   & 3.58    & 6.29        & 90.15           & $P2_1/m$                          \\
\hline
$\delta$               & 3.53    & 9.74* & 11.31     &-            & $Cmcm$                        \\
Exp. $\delta$\cite{Bouloux1976}               & 3.70 & 9.97*   & 11.02        & -           & $Cmcm$                          \\
\hline
$\gamma$                   & 9.88 & 3.56 & 10.05*      & -        & $Pnma$                          \\
Exp. $\gamma$\cite{Cocciantelli1995,Baddour-Hadjean2019}                 & 9.95 & 3.59 & 10.04*      & -           & $Pnma$                          \\
\hline
$\delta$-Ag$_{0.84}$   & 11.68  & 3.53 & 8.75*  & 91.03         & $C2/m$                         \\
Exp. $\delta$-Ag$_{0.84}$\cite{Rozier2009}              & 11.77    & 3.67 & 8.74*     & 90.54          & $C2/m$                          \\
\hline
$\epsilon$-Cu$_{0.85}$ & 11.5   & 3.58  & 9.01*  & 108.78           & $C2/m$                          \\
Exp. $\epsilon$-Cu$_{0.85}$\cite{Smirnov2018}               & 11.7    & 3.63 & 8.84*      & 109.6           & $C2/m$                          \\
\hline
$\nu$-Ca$_{0.6}$       & 11.52 & 3.58 & 9.50*  & 111.65    & $C2/m$                           \\
Exp. $\nu$-Ca$_{0.6}$\cite{Galy1992} & 11.81   & 3.71 & 9.27*      & 101.9           & $C2/m$                          \\
\hline
$\rho$-K$_{0.5}$       & 11.49 & 3.58 & 18.63*  & -     & $Cmcm$\\
Exp. $\rho$-K$_{0.5}$\cite{Galy1992}               & 11.61    & 3.67 & 18.67*      & -           & $Ccmm$                          \\
\hline
\end{tabular}
\label{lattConstTable}
\end{table}

\section{Electronic Properties} \label{sec:electronic}
Next, we turn our attention to the electronic structure of these polymorphs, where we focus on the single-layer $\alpha$ and the double-layer $\epsilon$-Cu$_{0.85}$ polymorph. Band structures corresponding to the other layered polymorphs are provided in the SI.

Table~\ref{bandsTable} lists the calculated direct and indirect band gaps, the location of the valence-band maximum (VBM) and conduction-band minimum (CBM), and the separation between the split-off bands and the other conduction bands. In all considered polymorphs the band gap is found to be indirect, with the VBM occurring on different high-symmetry lines, as indicated in the table. For the ground state $\alpha$ polymorph we find an indirect gap of 3.18 eV and a direct gap of 3.78 eV. 
Experimentally, the direct optical band gap for the $\alpha$ polymorph was measured to be 2.35 eV~\cite{Kenny1966}.
Calculations at the quasiparticle self-consistent GW (QSGW) level obtain indirect band gaps of 3.8-4.0 eV and direct gaps at $\Gamma$ of 4.4-4.8 eV~\cite{Bhandari2015,Gorelov2022,Garcia2024}. The discrepancy with the measured optical gap is attributed to excitons~\cite{Gorelov2022,Garcia2024}. Our HSE06 results underestimate the fundamental band gaps by 0.7 eV. The mixing parameter has to be increased to 35\% to obtain a better match. A test calculation for the double-layer $\epsilon$-Cu$_{0.85}$ polymorph, at 35\%, shows that the main effect of changing the mixing parameter is an increase of both the direct and indirect band gaps by 0.7 eV. Due to this systematic underestimation of the band gaps we opted to keep the default mixing parameter.

\begin{figure}[htb]
    \centering
    \includegraphics[width=0.7\columnwidth]{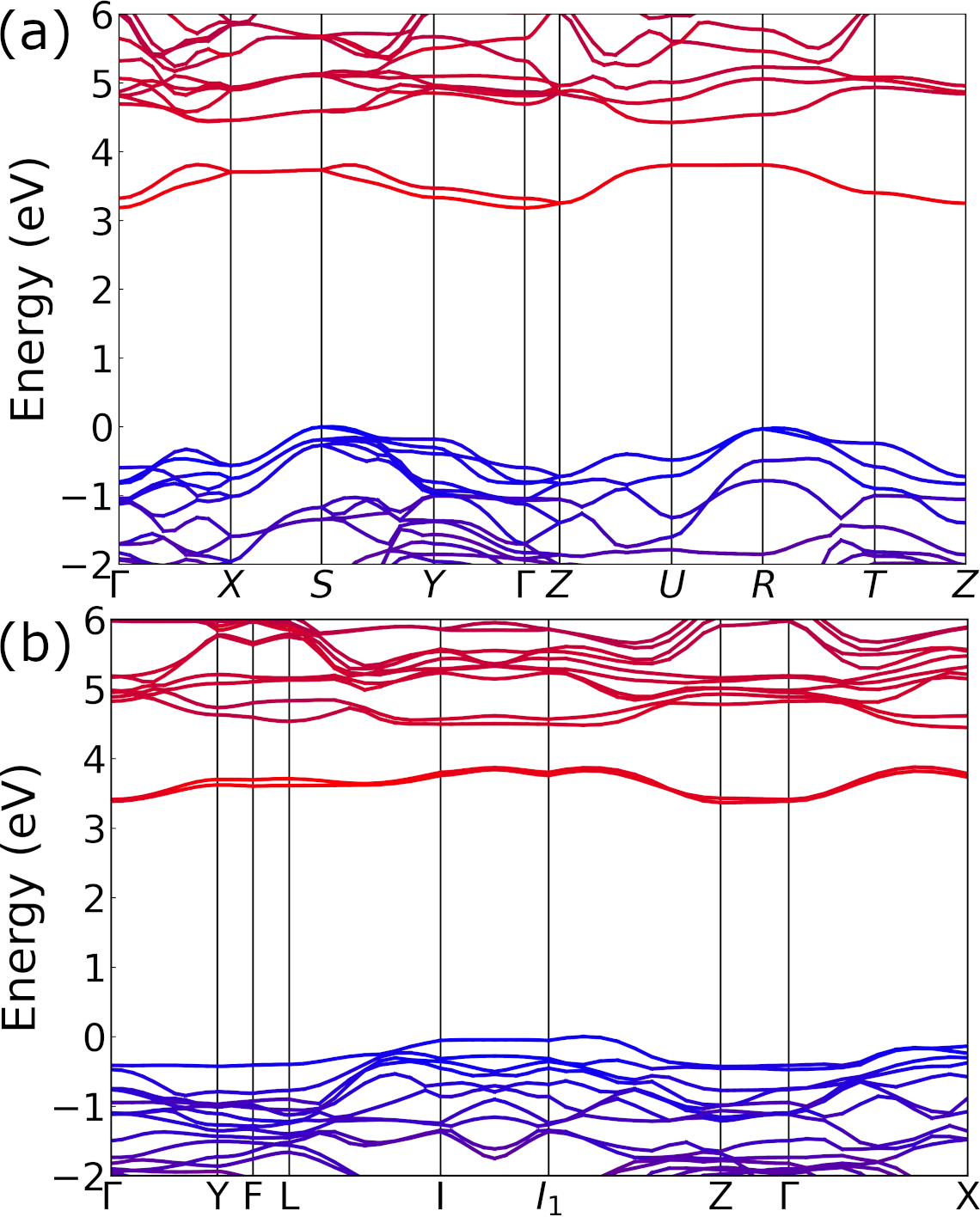}
    \caption{ The calculated band structures for the (a) single-layer $\alpha$-V$_2$O$_5$ polymorph and (b) the double-layer $\epsilon$-Cu$_{0.85}$ polymorph. The colors represent the band character, where red colors indicate V $d$ character and blue O $p$ character.}
    \label{AEbands}
\end{figure}

The resulting band structures of single- and double-layer polymorphs are remarkably similar, as illustrated in Fig.~\ref{AEbands}. The color scale corresponds to the orbital decomposition of the bands, where blue indicates O $p$ character and red V $d$ character. 
The highest valence bands are composed primarily of O $2p$ orbitals, while the lowest conduction bands are primarily V $3d$. Orbital mixing occurs for lower valence or higher conduction bands. 
All polymorphs exhibit split-off bands that are $\sim$0.6 eV below the other conduction bands (see Table~\ref{bandsTable}). The split-off bands originate from V-$d$ orbitals that have no antibonding interaction with the bridge oxygens (which link chains of V$_2$O$_5$ into layers)~\cite{Bhandari2015}.
However, in the high-temperature and high-pressure $\beta$ polymorph the split-off band merges with the other conduction bands on the high-symmetry $\Gamma$-$Z$ line. Band structures corresponding to the other single- and double-layered polymorphs are shown in the SI (Figs. S1-8). 
Due to the similar band characters for valence and conduction bands, all polymorphs have similar direct and indirect band gaps, with the exception of the high-temperature and high-pressure $\beta$ polymorph. The electronic properties of all polymorphs are therefore extremely similar.

\begin{table}[]
\caption{Location of the VBM and CBM, magnitude of the indirect and direct band gap at the $\Gamma$ point, and the energy separation of the split-off band.}

\begin{tabular}{lccccc}
\multicolumn{1}{r}{} & VBM                          & CBM   & $E^{\rm indirect}_{\rm gap}$  & $E^{\rm direct}_{\rm gap}$ & $E^{\rm split-off}$\\
\hline
$\alpha$               & S $\longrightarrow$ Y        & $\Gamma$ & 3.18              & 3.78                       & 0.61                          \\
$\beta$                & C $\longrightarrow$ E        & Y        & 2.74              & 3.06                       & ---          \\
$\delta$               & $A_1$ $\longrightarrow$ T    & $\Gamma$ & 3.69              & 3.84                       & 0.60                         \\
$\gamma$               & Z $\longrightarrow$ U        & $\Gamma$ & 3.38              & 3.55                       & 0.58                          \\
$\delta$-Ag$_{0.84}$   & $I_1$ $\longrightarrow$ Z    & $\Gamma$ & 3.47              & 3.8                       & 0.6                          \\
$\epsilon$-Cu$_{0.85}$ & $I_1$ $\longrightarrow$ Z    & Z        & 3.37              & 3.8                        & 0.57                          \\
$\nu$-Ca$_{0.6}$       & $\Gamma$ $\longrightarrow$ X & $\Gamma$ & 3.45              & 3.79                       & 0.6                           \\
$\rho$-K$_{0.5}$       & $\Gamma$ $\longrightarrow$ X & $\Gamma$ & 3.46              & 3.79                       & 0.58                         
\end{tabular}
\label{bandsTable}
\end{table}

\section{Electronic role of intercalants}\label{sec:intercalated}

Thus far, we focused on unintercalated polymorphs. We will consider the role of intercalants on the electronic properties next.
The role of intercalants in the $\alpha$ V$_2$O$_5$ polymorph was previously studied computationally for intercalation with Li~\cite{Watthaisong2019}, Na ~\cite{Bhandari2015}, and both Li and Mg~\cite{Xiao2018}. Intercalation in the $\gamma$ polymorph was studied for Li and Na~\cite{Roginskii2021}. 
Consistently, these works show that the intercalant orbitals do not lead to states within the band gap region. The band gap region itself does contain bands with V-d character, but these correspond to lowered existing conduction band states. ~\cite{Watthaisong2019}.
These results are therefore consistent with our hypothesis that the primary electronic effect of intercalants is the formation of V-d character midgap states through the donation of electrons. Structurally, intercalants can stabilize phases, as evident from the experimental evidence detailed in the introduction.

To further elucidate the electronic effects of intercalants, we investigated the $\alpha$ single-layer and $\epsilon$-Cu$_{0.85}$ bilayer polymorphs intercalated with 0.5 Li per formula unit. %
Information about more concentrations and other intercalants (Mg, Zn, K) can be found in the SI. For Na intercalation in $\alpha$-V$_2$O$_5$ it was found that the magnetic moments on V-sites along the chain direction are antiferromagnetically (AFM) ordered~\cite{Bhandari2015}.
We therefore explored both non-magnetic and AFM orderings in the $\alpha$ polymorph and found that for all concentrations and considered intercalants (Li$^+$, K$^+$, Mg$^{2+}$, and Zn$^{2+}$) that the AFM ordering is indeed the most favorable ordering (by $\sim$15 meV per formula unit for Li, Mg, and Zn, and by 40 meV for K). In contrast, unintercalated $\alpha$ prefers a non-magnetic ordering.

\begin{figure}[htb]
    \centering
    \includegraphics[width=0.7\columnwidth]{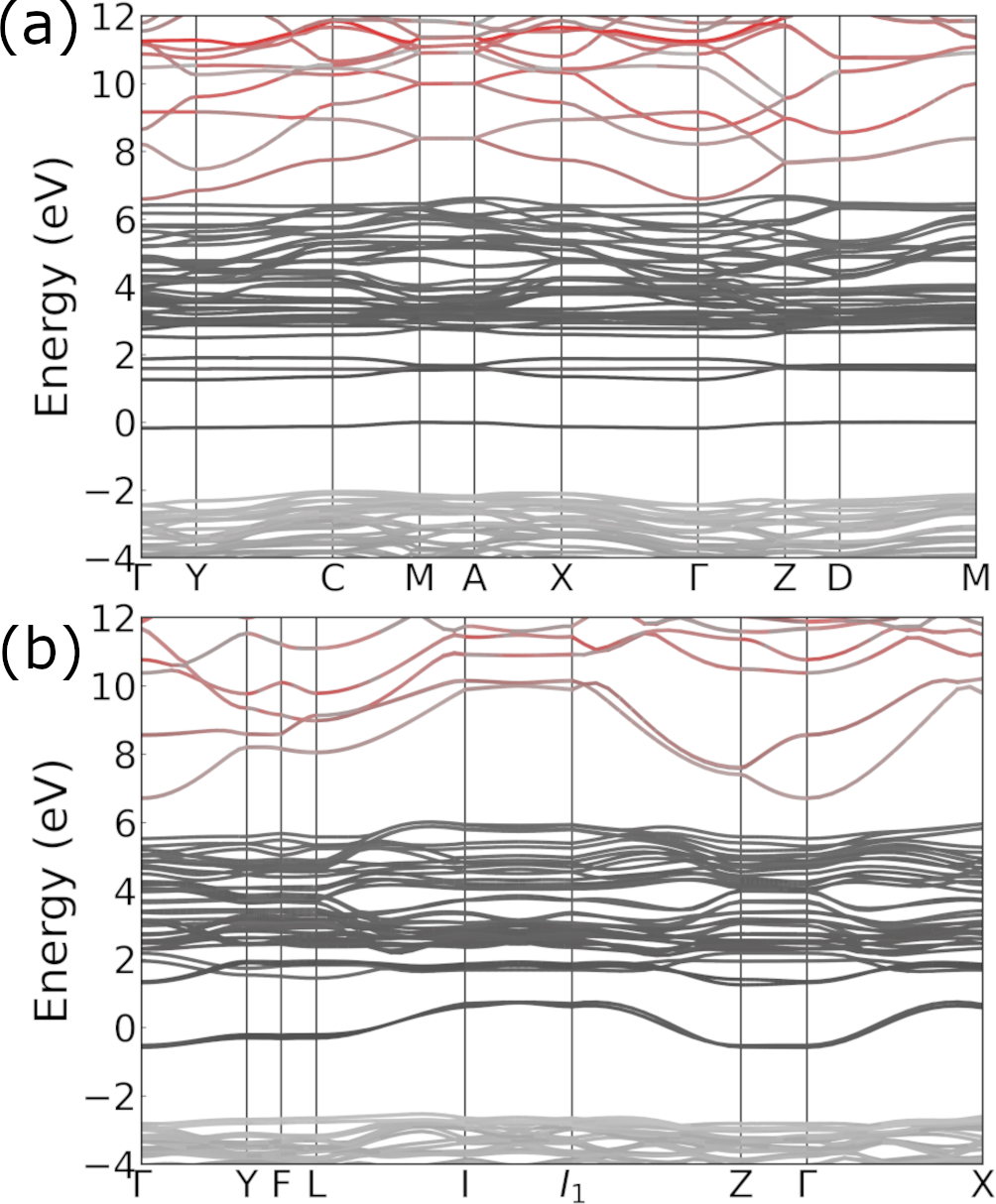}
    \caption{ The calculated band structures for a 0.5 Li per formula unit intercalation of the (a) single-layer $\alpha$-V$_2$O$_5$ polymorph with AFM ordering and (b) the double-layer $\epsilon$-Cu$_{0.85}$ polymorph. The V contribution is colored dark grey, the O contribution in light grey, and the Li contribution in red. The Fermi level is used as the energy 0 reference.}
    \label{LiAEbands}
\end{figure}

In Fig.~\ref{LiAEbands}, we show the band structures for Li$_{0.5}$V$_2$O$_5$ in both the single-layer $\alpha$ (with AFM) and double-layer $\epsilon$-Cu$_{0.85}$ polymorphs. We color coded the band structures so that V contributions are depicted in dark grey, O contributions in light grey, and Li contributions in red. These band structures immediately reveal that Li contributions are located high in the conduction bands (more than 5 eV above the CBM). Similarly for Mg$_{0.5}$ (Figs. S13-S14), K$_{0.5}$ (Figs. S15-S16), and Zn$_{0.5}$ (Figs. S17-S18), the intercalant only contributes high above the CBM. Note that since Zn also has d orbitals, some small Zn contributions show up in the valence bands and its conduction band contribution is lower compared to the other intercalants (but still 4 eV above the CBM). We tested increased intercalant concentrations (that maintain the layered structure, e.g., Li$_1$ shown in Fig. S11). The increase in intercalant concentration results in more available electrons, so that more of the split-off bands become occupied. The number of newly occupied bands only depends on the introduced carrier concentration, e.g., Li$_1$ (Fig. S11) leads to the same number of occupied bands as Mg$_{0.5}$ (Fig. S13). The shape of these newly filled bands does depend slightly on the intercalants and their concentration, due to small relaxation effects of the underlying V$_2$O$_5$ structure.

Since all intercalant states are located well above the CBM, the corresponding electrons will fill the lowest conduction-band states, i.e., those of the split-off bands. Once electrons occupy conduction bands, their band energies decrease, leading to a separation from the other conduction bands. Note that the now filled bands are either completely filled (as is the case for the $\alpha$ polymorph) or multiple split-off bands are partially filled (for the $\epsilon$-Cu$_{0.85}$ polymorph).

\section{Conclusions}
In conclusion, we performed a detailed first-principles study of 4 single-layer and 4 double-layer unintercalated V$_2$O$_5$ polymorphs using hybrid density functional theory. We compared several methods to include van der Waals interactions and found that the Grimme D3 method provided the most accurate results. Our band structure calculations reveal that independent of the studied polymorph, the band structure has very similar features: the topmost valence bands mainly consist of O $p$ character, while the lowest conduction bands consist of V $d$ orbitals. The lowest conduction bands are well separated from the other conduction bands, and the energy difference between these bands and the next conduction bands is around 0.6 eV for all polymorphs. The band gaps are also similar. Only the high-temperature and high-pressure $\beta$ polymorph has a band structure where the split-off band merges with the other conduction bands for some of the high-symmetry directions, and a smaller band gap. We determined the main electronic role of intercalants for the $\alpha$ and $\epsilon$-Cu$_{0.85}$ polymorphs. The contributions of Li$^+$, K$^+$, Mg$^{2+}$, and Zn$^{2+}$ are located well above the conduction-band minimum. Their main role is therefore to donate electrons that will occupy the split-off bands, and upon occupation, lower the energy of these bands. 
Our results show that the electronic properties of various layered V$_2$O$_5$ polymorphs are robust across all polymorphs and even when intercalants are present. This robustness will therefore also translate to V$_2$O$_5$-based cathodes upon cycling of batteries.

\section*{Acknowledgements}
This work was supported by the National Science Foundation (NSF) through DMR-2339751 and by the University of Kansas General Research Fund allocation \#2151089.
This research used resources of the National Energy Research
Scientific Computing Center, a DOE Office of Science User Facility
supported by the Office of Science of the U.S. Department of Energy
under Contract No. DE-AC02-05CH11231 using NERSC award
BES-ERCAP0033524.

\section{Data availability}
The data supporting this article have been included as part of the Supplementary Information (SI). The SI contains tabular data (Table S1 and S2) of lattice constants and percentage deviations from experiment of the single layer $\alpha$, $\beta$, and nonlayered B polymorphs with respect to the various van der Waals interaction methods. Fig. S1 - S8 show the electronic band structures of the unintercalated polymorphs. Fig S9 - S18 show the electronic band structures for the $\alpha$ and $\epsilon$-Cu$_{0.85}$ polymorphs with Li, Mg, K and Zn intercalants. A zip file of our relaxed structures used for band structure calculations, in VASP POSCAR format, are also included.

\bibliography{main}

\clearpage
\appendix
\section*{Supplementary Information}

\section{Lattice constants}
This section has tables, S1 and S2, detailing the lattice constants of the $\alpha$, $\beta$ and non-layered B polymorphs as obtained with HSE06 in combination with various approaches to include van der Waals interactions.

\section{Unintercalated Bandstructures}
Bandstructures of the unintercalated layered polymorphs of V$_2$O$_5$ are shown in Figs. S1-S8. The color scale indicates the bands' orbital character, where red indicates vanadium and blue oxygen character. All band structures were calculated with HSE06 with Grimme D3 to describe the van der Waals interactions, as described in the main manuscript.

\section{Intercalated Bandstructures}
Bandstructures of the intercalated $\alpha$ and $\epsilon$-Cu$_{0.85}$ polymorphs with: Li at two concentrations (Fig S9-12), Mg (Fig. S13-14), K (Fig. S15-16) and Zn (Fig. S17-18) as described in their captions. Dark grey bands are V-character, light grey bands are O-character, and colored bands are intercalant character. Dashed and solid lines represent the two spins.

\setcounter{table}{0}  
\renewcommand{\thetable}{S\arabic{table}}  
\setcounter{figure}{0}
\renewcommand{\thefigure}{S\arabic{figure}}

\begin{table}[h]
\caption{Lattice parameters and percentage deviations for the $\alpha$ and $\beta$ polymorphs using different van der Waals; interactions.}
\begin{tabular}{llcccccccc}
polymorph &
   &
  a (\AA) &
  b (\AA) &
  c (\AA) &
  $\beta$ (deg.) &
  $\Delta$ a (\%) &
  $\Delta$ b (\%) &
  $\Delta$ c (\%) &
  $\Delta$ $\beta$ (\%) \\\hline\hline
 &
  experimental (Ref. 66) &
  11.512 &
  3.564 &
  4.368* &
   &
   &
   &
   &
   \\
 &
  HSE06 &
  11.362 &
  3.545 &
  4.689* &
   &
  -1.30 &
  -0.53 &
  7.35* &
   \\
 &
  HSE-D2 &
  11.438 &
  3.512 &
  4.415* &
   &
  -0.64 &
  -1.46 &
  1.08* &
   \\
 &
  HSE-D3+damping &
  11.428 &
  3.527 &
  4.331* &
   &
  -0.73 &
  -1.04 &
  -0.86* &
   \\
$\alpha$ &
  HSE-D3 &
  11.434 &
  3.532 &
  4.363* &
   &
  -0.67 &
  -0.90 &
  -0.11* &
   \\
 &
  HSE-D4 &
  11.431 &
  3.527 &
  4.359* &
   &
  -0.70 &
  -1.04 &
  -0.21* &
   \\
 &
  HSE-TS &
  11.448 &
  3.521 &
  4.347* &
   &
  -0.56 &
  -1.19 &
  -0.49* &
   \\
 &
  HSE-TSH &
  11.458 &
  3.518 &
  4.337* &
   &
  -0.47 &
  -1.30 &
  -0.71* &
   \\
 &
  HSE-rVV10 &
  11.479 &
  3.528 &
  4.215* &
   &
  -0.29 &
  -1.01 &
  -3.51* &
   \\
 &
HSE-rVV10L &
  11.450 &
  3.533 &
  4.339* &
   &
  -0.53 &
  -0.86 &
  -0.66* &
   \\
 &
   &
   &
   &
   &
   &
   &
   &
   &
   \\
 &
  experimental (Ref. 39) &
  7.112* &
  3.579 &
  6.290 &
  90.15 &
   &
   &
   &
   \\
 &
  HSE06 &
  7.863* &
  3.539 &
  6.278 &
  91.14 &
  10.56* &
  -1.13 &
  -0.19 &
  1.10 \\
 &
  HSE-D2 &
  7.092* &
  3.526 &
  6.237 &
  90.48 &
  -0.28* &
  -1.49 &
  -0.85 &
  0.37 \\
 &
  HSE-D3+damping &
  7.079* &
  3.535 &
  6.231 &
  90.40 &
  -0.46* &
  -1.24 &
  -0.95 &
  0.27 \\
$\beta$ &
  HSE-D3 &
  7.119* &
  3.529 &
  6.261 &
  90.39 &
  0.10* &
  -1.41 &
  -0.47 &
  0.27 \\
 &
  HSE-D4 &
  7.149* &
  3.523 &
  6.245 &
  90.28 &
  0.52* &
  -1.57 &
  -0.73 &
  0.15 \\
 &
  HSE-TS &
  7.066* &
  3.530 &
  6.249 &
  90.07 &
  -0.64* &
  -1.38 &
  -0.65 &
  -0.08 \\
 &
  HSE-TSH &
  7.145* &
  3.529 &
  6.236 &
  90.11 &
  0.47* &
  -1.41 &
  -0.86 &
  -0.05 \\
 &
  HSE-rVV10 &
  6.990* &
  3.533 &
  6.213 &
  90.428 &
  -1.71* &
  -1.30 &
  -1.22 &
  0.31 \\
 &
  HSE-rVV10L &
  7.153* &
  3.532 &
  6.250 &
  90.385 &
  0.57* &
  -1.31 &
  -0.65 &
  0.26
\end{tabular}
\end{table}

\begin{table}[h]
\caption{Lattice parameters and percentage deviations for the B polymorph using different van der Waals interactions.}
\begin{tabular}{llcccccccc}
polymorph &
   &
  a (\AA) &
  b (\AA) &
  c (\AA) &
  $\beta$ (deg.) &
  $\Delta$ a (\%) &
  $\Delta$ b (\%) &
  $\Delta$ c (\%) &
  $\Delta$ $\beta$ (\%) \\ \hline \hline
 &
  experimental (Ref. 39) &
  11.972 &
  4.702 &
  5.325 &
  104.41 &
   &
   &
   &
   \\
             & HSE06      & 11.801 & 4.644 & 5.361 & 103.60 & -1.43 & -1.24 & 0.68  & -0.78 \\
             & HSE-D2         & 11.761 & 4.619 & 5.296 & 104.20 & -1.77 & -1.78 & -0.54 & -0.20 \\
B            & HSE-D3+damping & 11.766 & 4.613 & 5.284 & 104.06 & -1.72 & -1.90 & -0.78 & -0.33 \\
(nonlayered) & HSE-D3         & 11.807 & 4.626 & 5.333 & 103.97 & -1.38 & -1.61 & 0.14  & -0.42 \\
             & HSE-D4         & 11.544	&4.673	&5.305 & 105.28 & -3.57	&-0.61	&-0.38	&0.83 \\
             & HSE-TS         & 11.798 & 4.603 & 5.293 & 103.92 & -1.46 & -2.10 & -0.60 & -0.47 \\
             & HSE-TSH        & 11.770 & 4.611 & 5.284 & 103.95 & -1.68 & -1.94 & -0.78 & -0.45 \\
             & HSE-rVV10      & 11.768 & 4.605 & 5.256 & 104.23 & -1.70 & -2.06 & -1.30 & -0.17 \\
             & HSE-rVV10L     & 11.784 & 4.623 & 5.302 & 103.94 & -1.57 & -1.68 & -0.43 & -0.45
\end{tabular}
\end{table}

\begin{figure}[p]
    \centering
    \includegraphics[width=0.8\columnwidth]{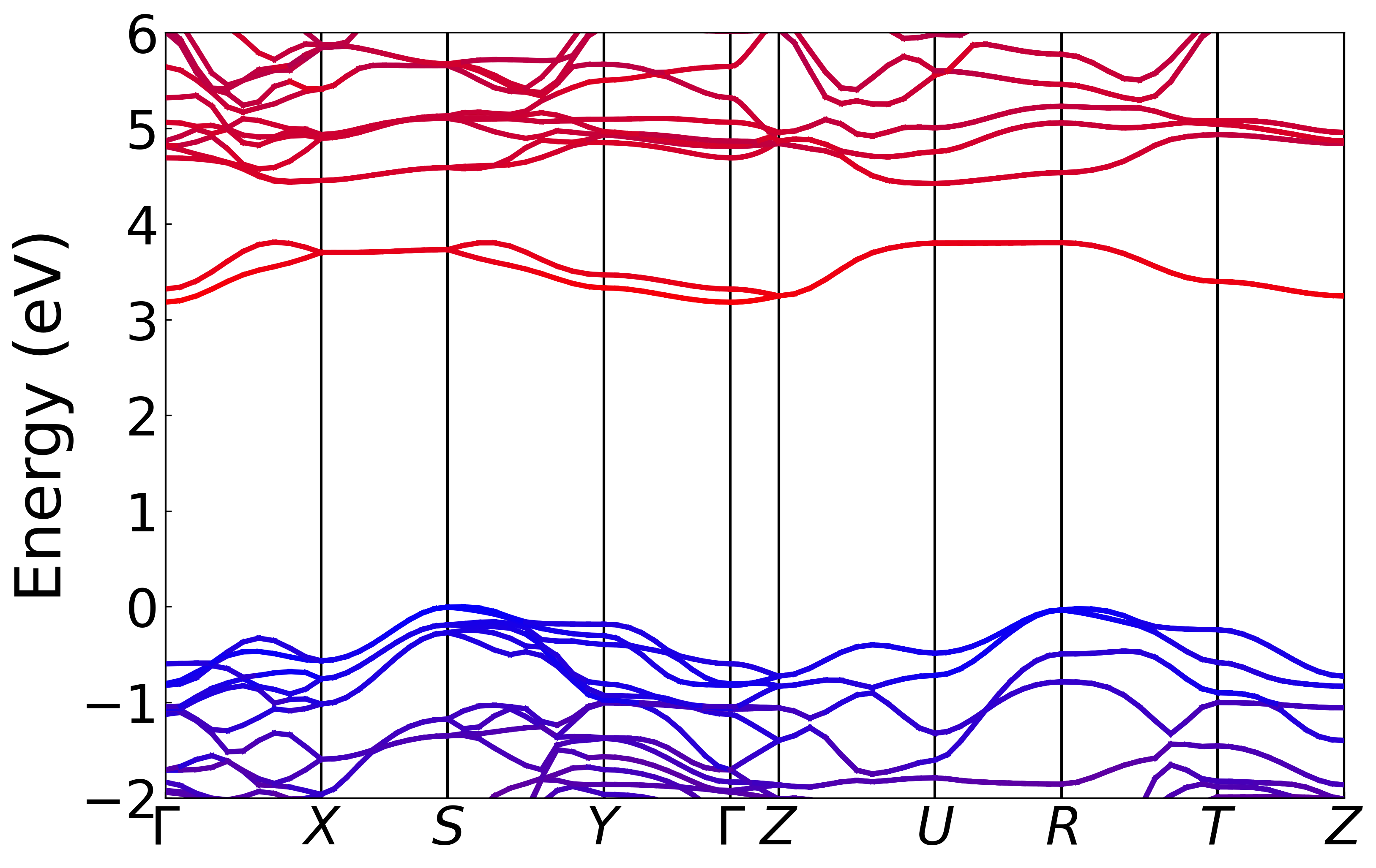}
    \caption{Calculated bandstructure for the $\alpha$ polymorph (delithiated Li$_x$V$_2$O$_5$ for $0 < x < 0.1$).}
    \label{alphaSing}
\end{figure}
\begin{figure}
    \centering
    \includegraphics[width=0.8\columnwidth]{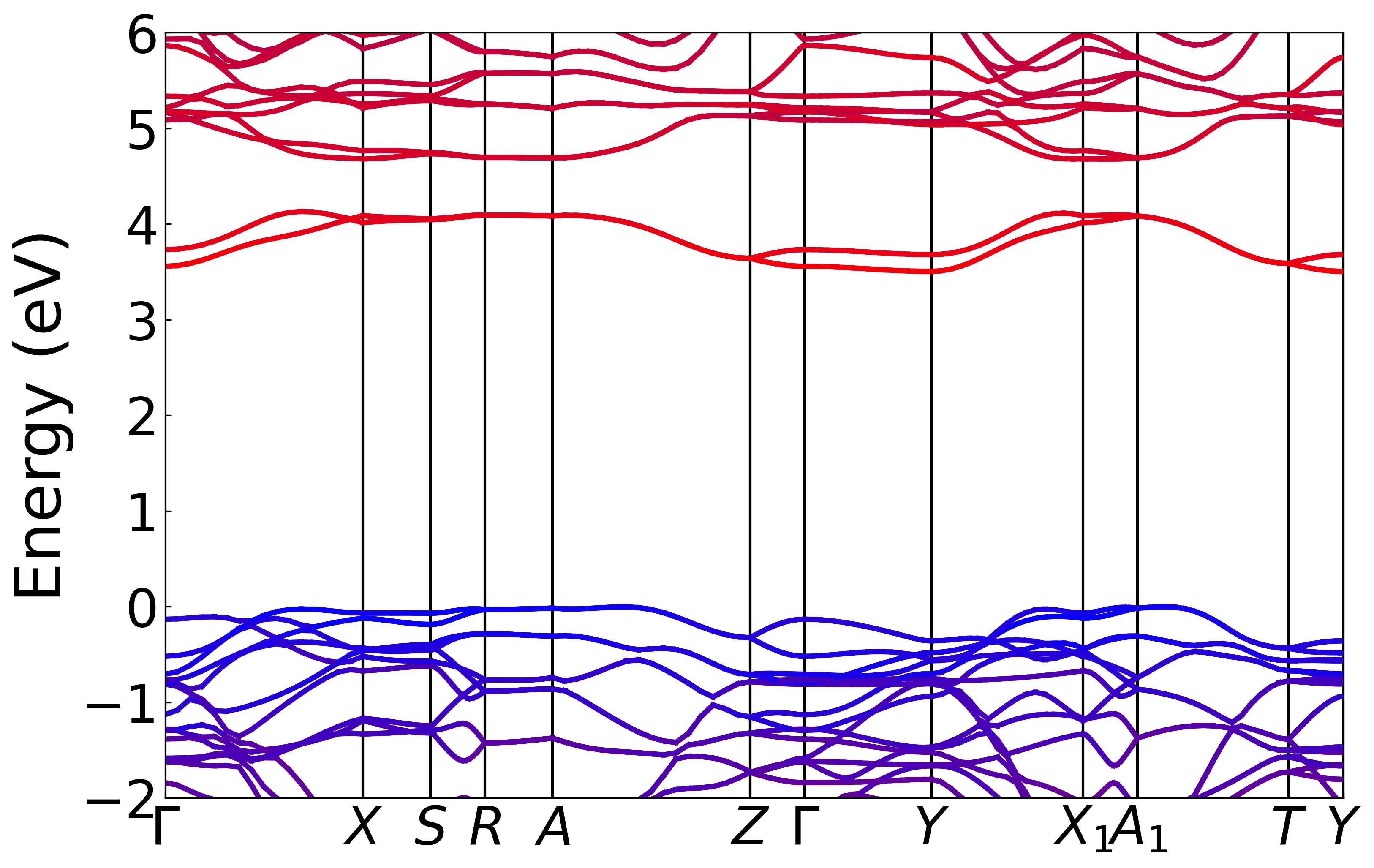}
    \caption{Calculated bandstructure for the $\delta$ polymorph (delithiated Li$_x$V$_2$O$_5$ for $0.7 < x < 1$).}
    \label{delSing}
\end{figure}
\begin{figure}
    \centering
    \includegraphics[width=0.8\columnwidth]{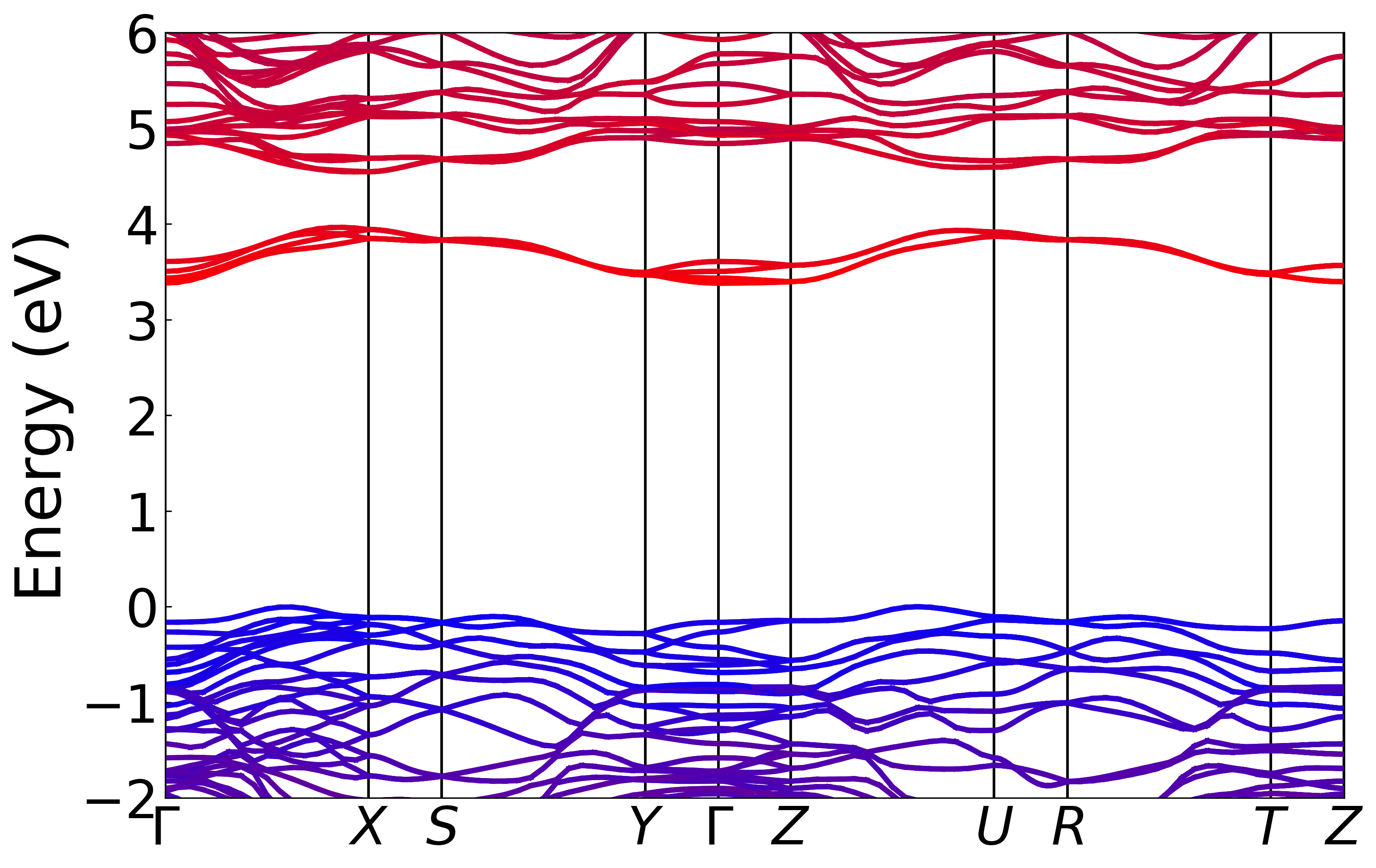}
    \caption{Calculated bandstructure for the $\gamma$ polymorph (delithiated Li$_x$V$_2$O$_5$ for $1 < x < 2$).}
    \label{gamSing}
\end{figure}
\begin{figure}
    \centering
    \includegraphics[width=0.8\columnwidth]{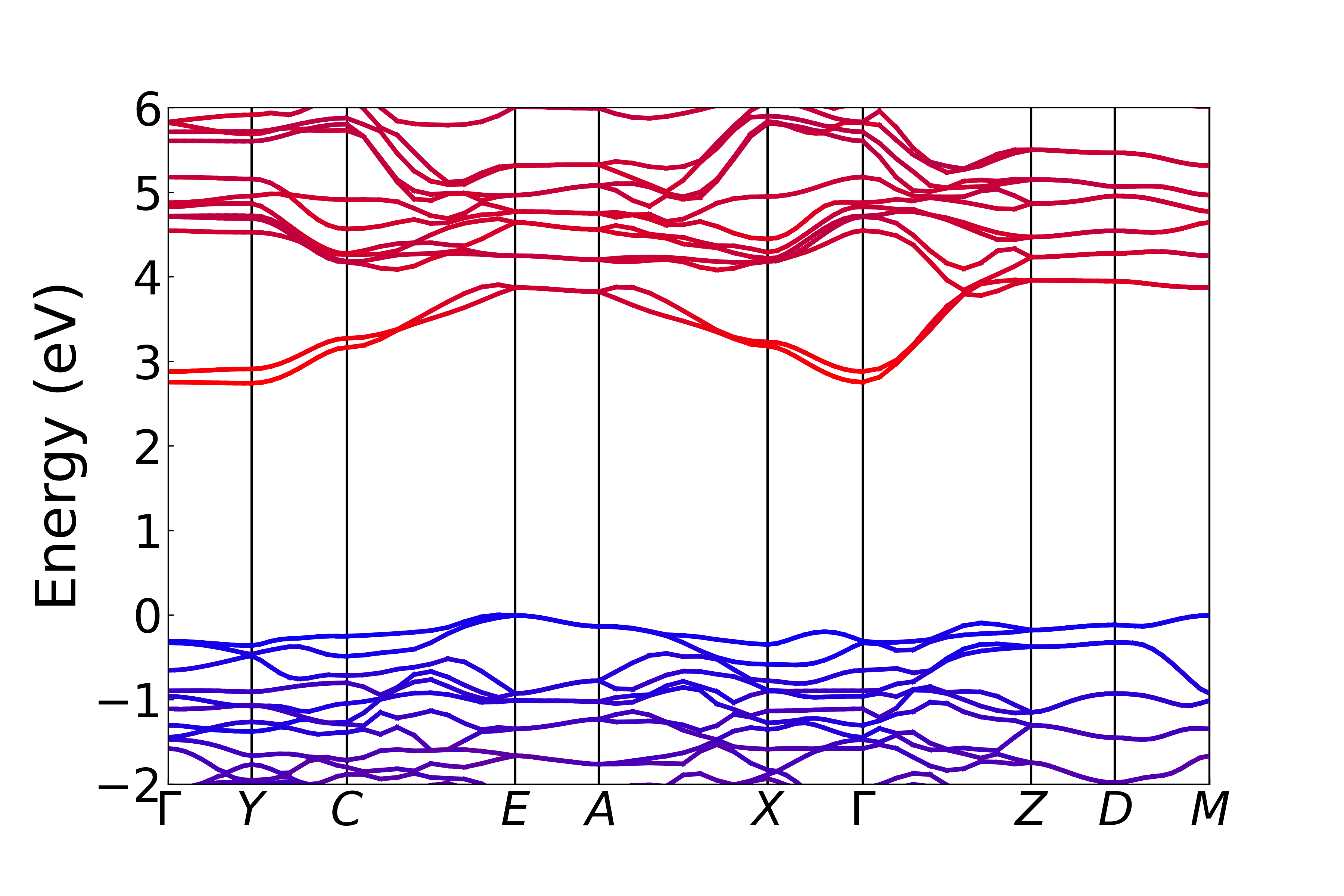}
    \caption{Calculated bandstructure for the $\beta$ polymorph (a high temperature and high pressure polymorph of V$_2$O$_5$).}
    \label{betSing}
\end{figure}
\begin{figure}
    \centering
    \includegraphics[width=0.8\columnwidth]{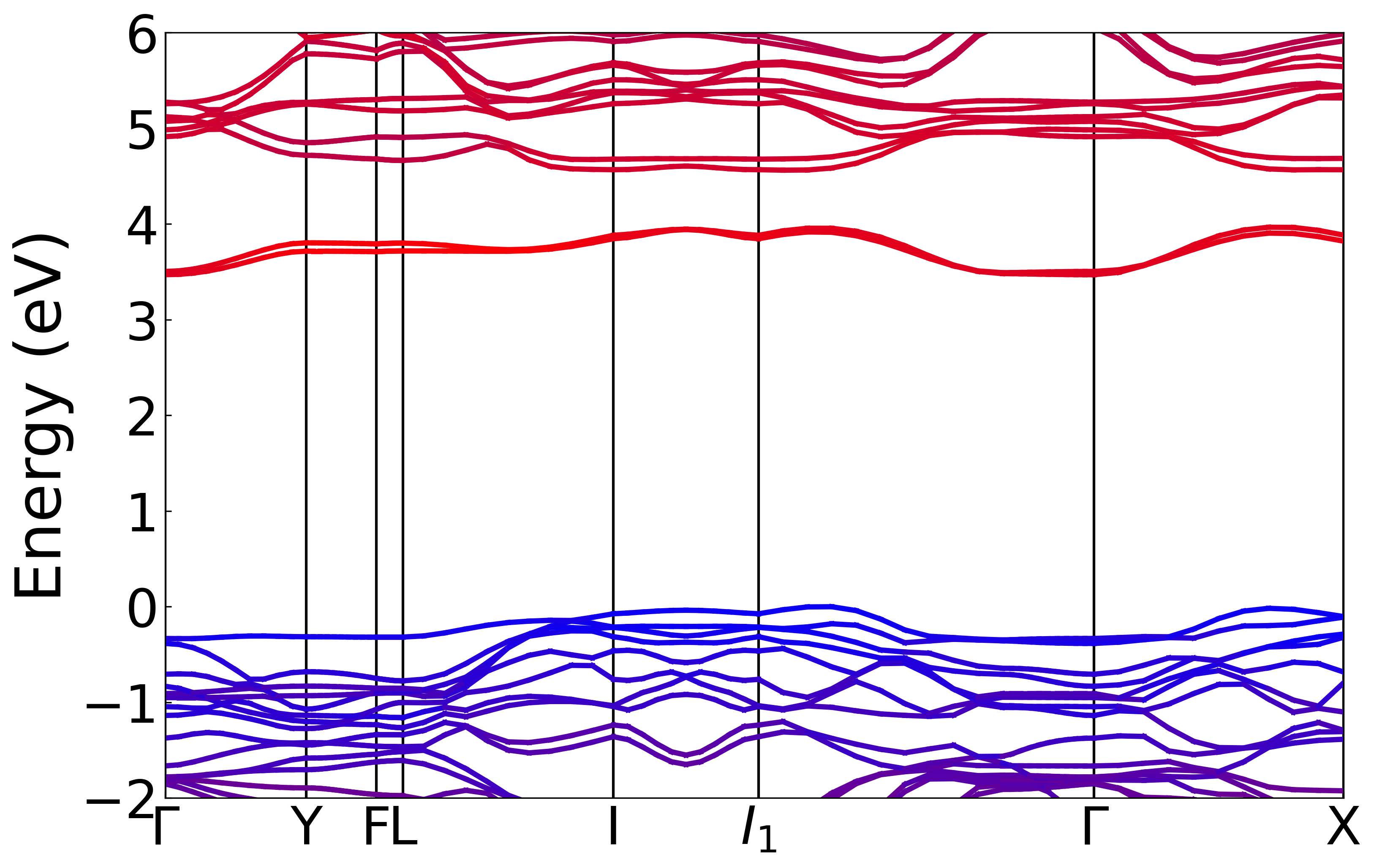}
    \caption{Calculated bandstructure for the double-layer $\delta$-Ag$_{0.84}$ polymorph (unintercalated Ag$_{0.84}$V$_2$O$_5$).}
    \label{deltDub}
\end{figure}
\begin{figure}
    \centering
    \includegraphics[width=0.8\columnwidth]{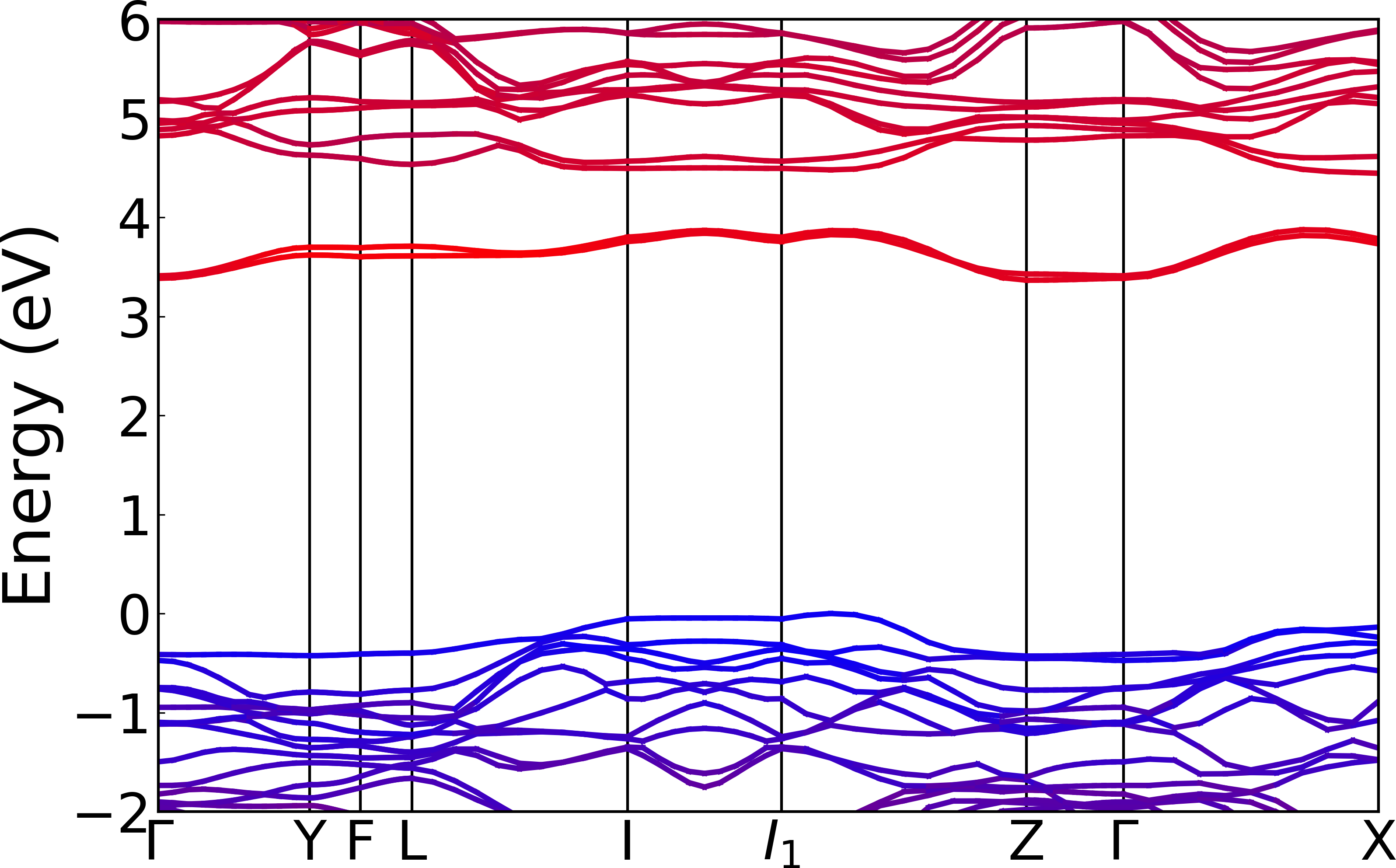}
    \caption{Calculated bandstructure for the double-layer $\epsilon$-Cu$_{0.85}$ polymorph (unintercalated Cu$_{0.85}$V$_2$O$_5$).}
    \label{epsDub}
\end{figure}
\begin{figure}
    \centering
    \includegraphics[width=0.8\columnwidth]{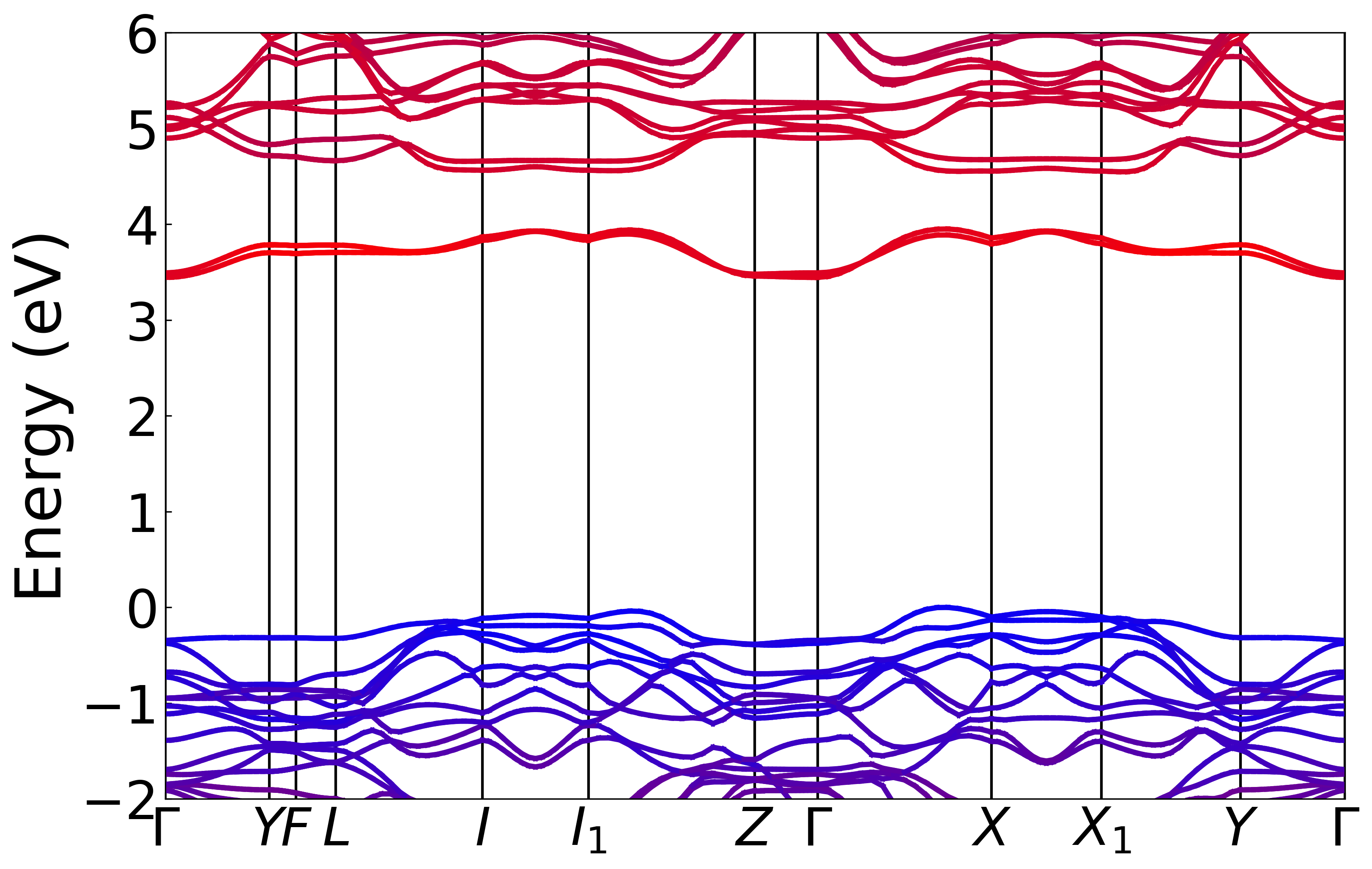}
    \caption{Calculated bandstructure for the double-layer $\nu$-Ca$_{0.6}$ polymorph (unintercalated Ca$_{0.6}$V$_2$O$_5$).}
    \label{nuDub}
\end{figure}
\begin{figure}
    \centering
    \includegraphics[width=0.8\columnwidth]{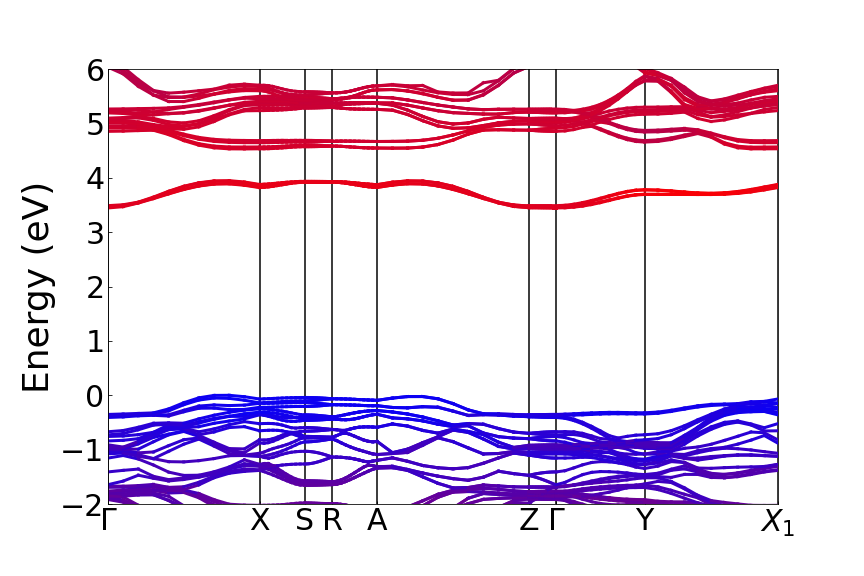}
    \caption{Calculated bandstructure for the double-layer $\rho$-K$_{0.5}$ polymorph (unintercalated K$_{0.5}$V$_2$O$_5$).}
    \label{rhoDub}
\end{figure}
\begin{figure}
    \centering
    \includegraphics[width=0.8\columnwidth]{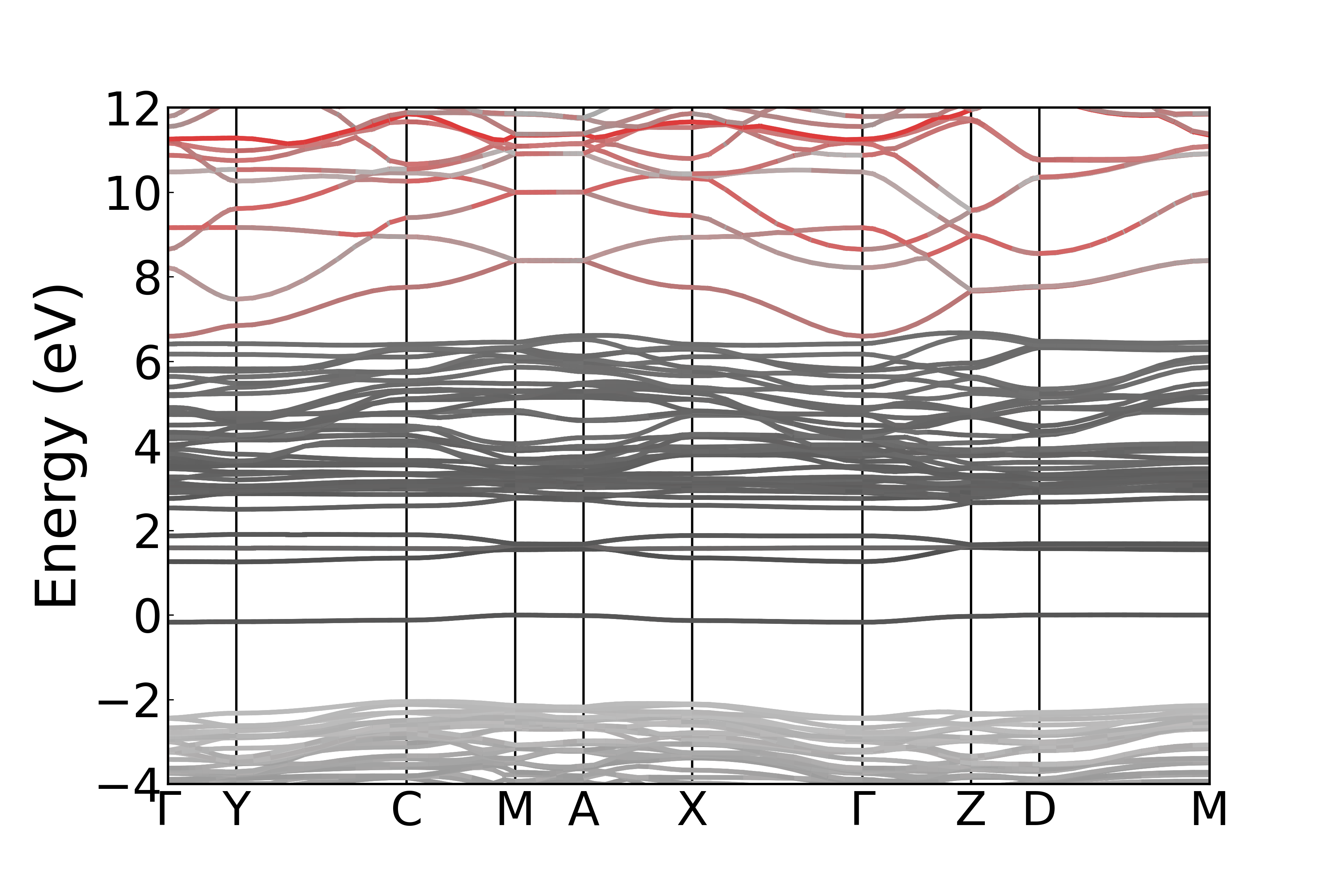}
    \caption{Calculated bandstructure of $\alpha$-Li$_{0.5}$V$_2$O$_5$ with AFM ordering.}
    \label{Li05alphaAFM}
\end{figure}
\begin{figure}
    \centering
    \includegraphics[width=0.8\columnwidth]{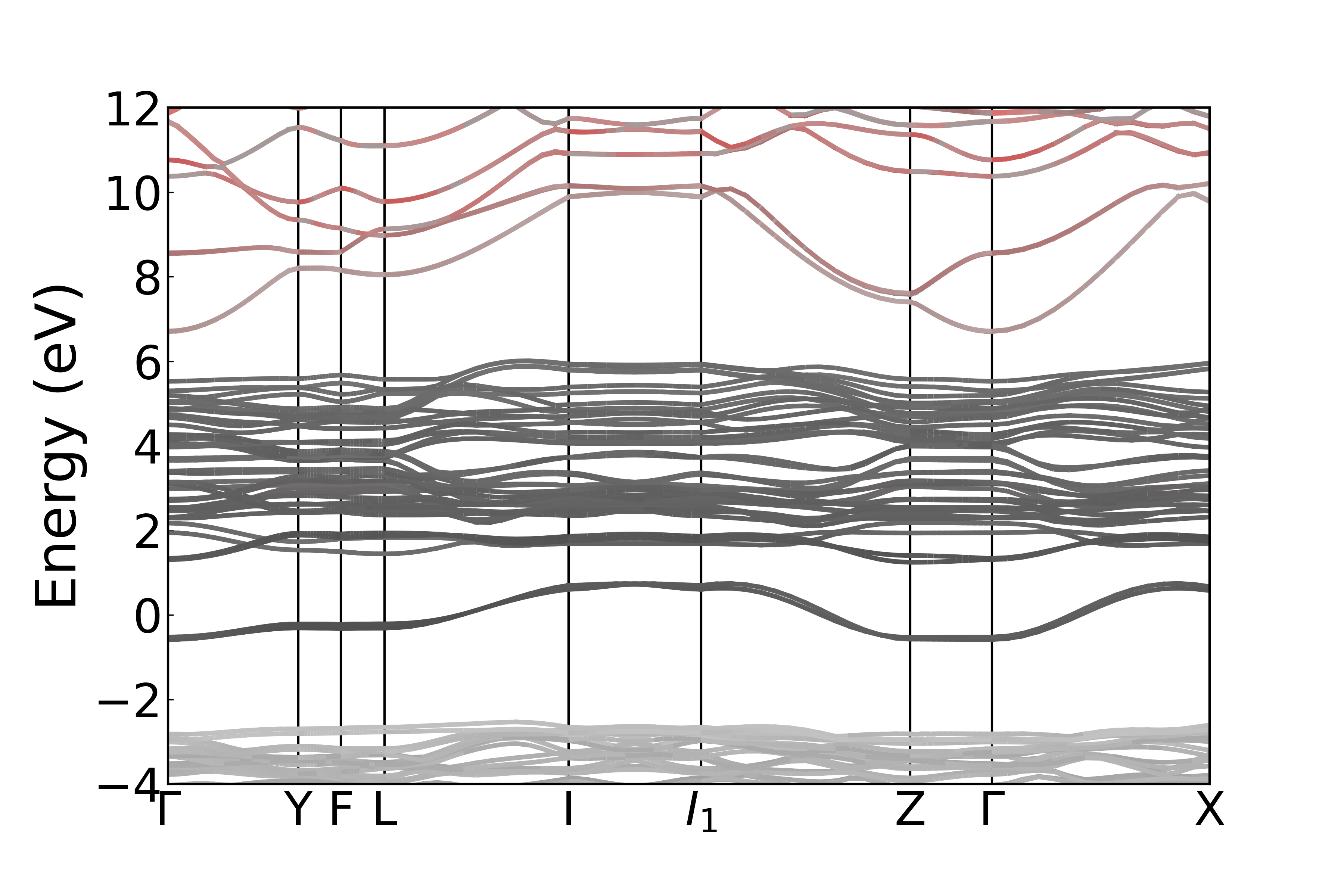}
    \caption{Calculated bandstructure of Li$_{0.5}$V$_2$O$_5$ in the $\epsilon$-Cu$_{0.85}$ polymorph.}
    \label{Li05eps}
\end{figure}
\begin{figure}
    \centering
    \includegraphics[width=0.8\columnwidth]{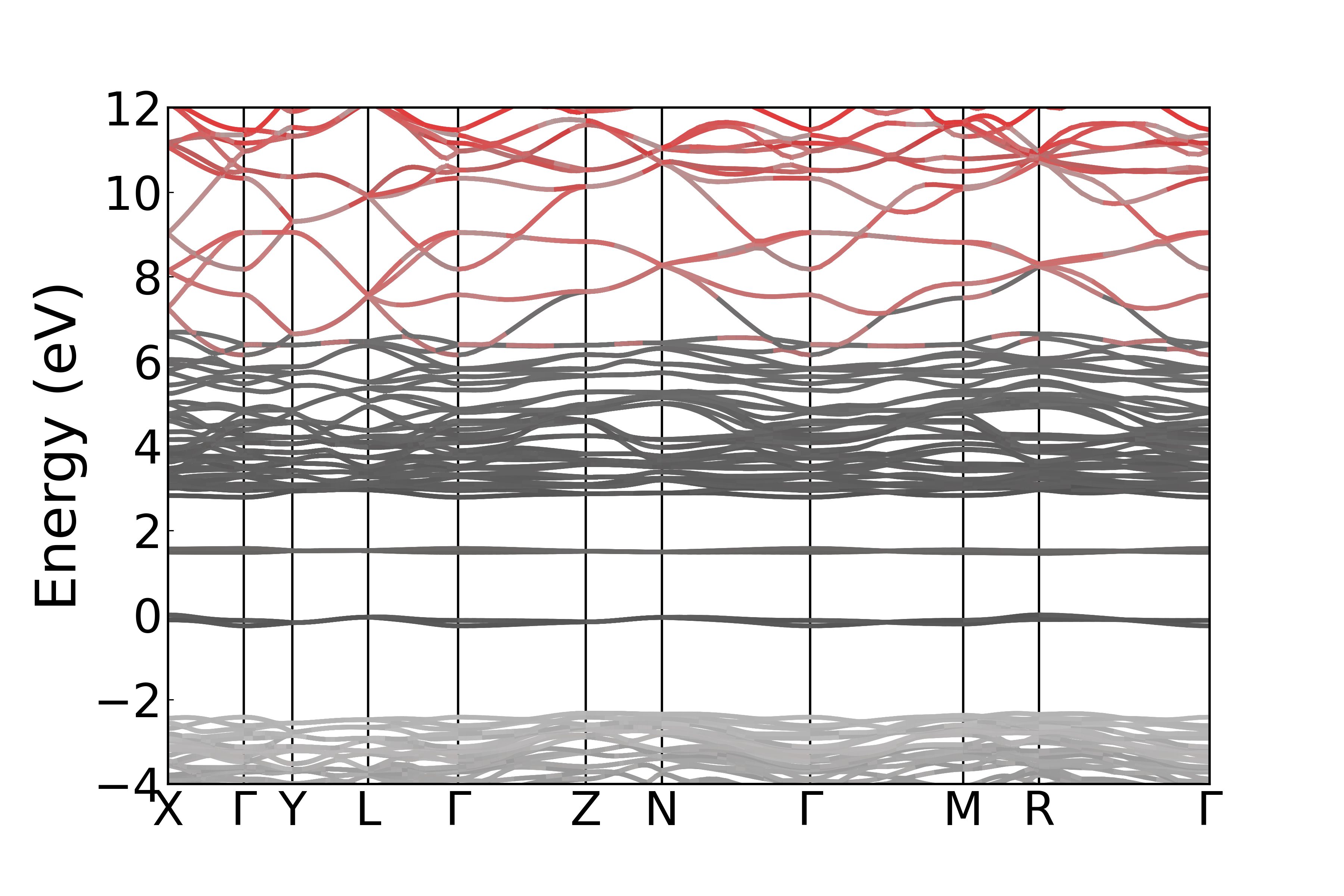}
    \caption{Calculated bandstructure of $\alpha$-Li$_1$V$_2$O$_5$ with AFM ordering.}
    \label{Li1alphaAFM}
\end{figure}
\begin{figure}
    \centering
    \includegraphics[width=0.8\columnwidth]{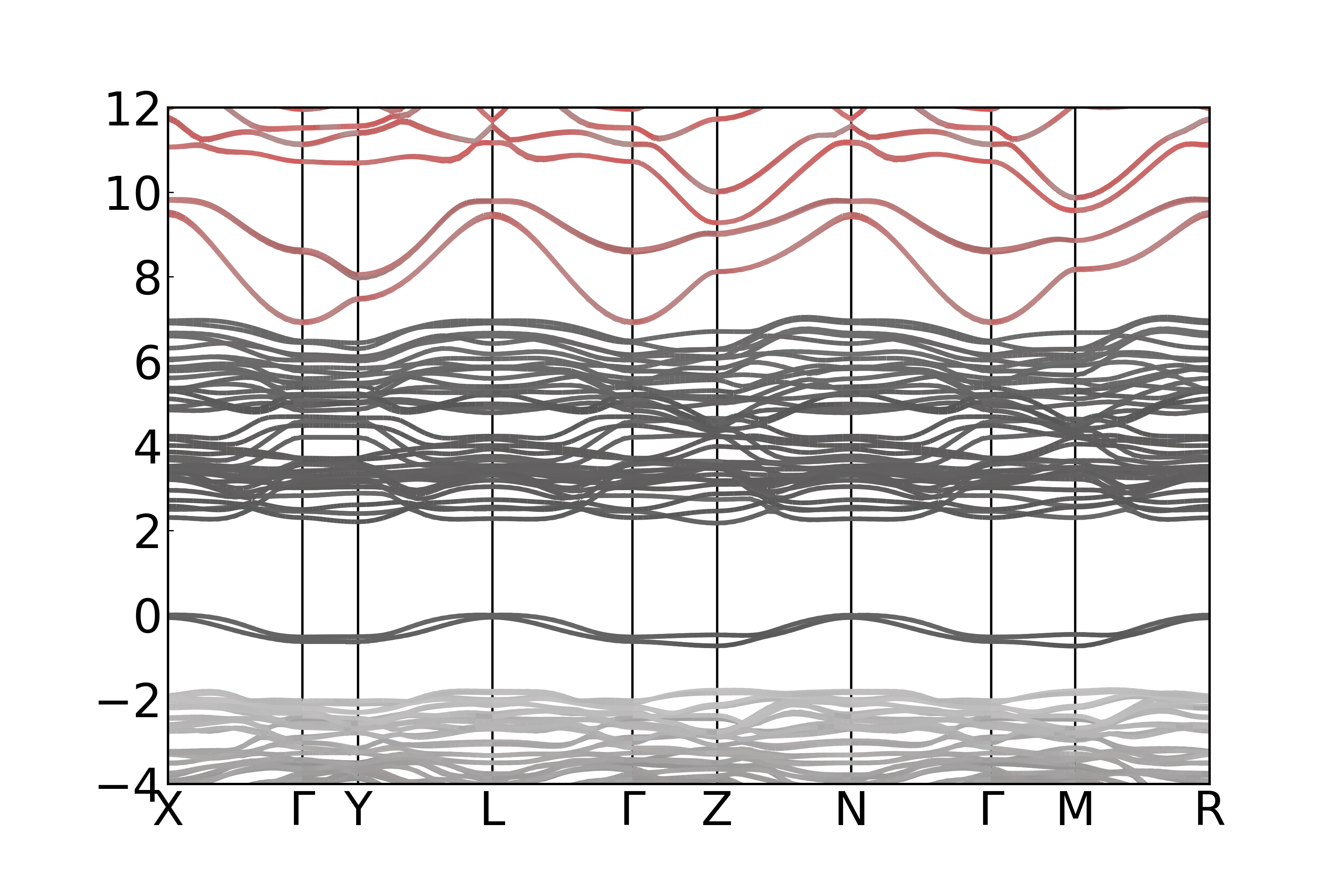}
    \caption{Calculated bandstructure of Li$_1$V$_2$O$_5$ in the $\epsilon$-Cu$_{0.85}$ polymorph.}
    \label{Li1eps}
\end{figure}
\begin{figure}
    \centering
    \includegraphics[width=0.8\columnwidth]{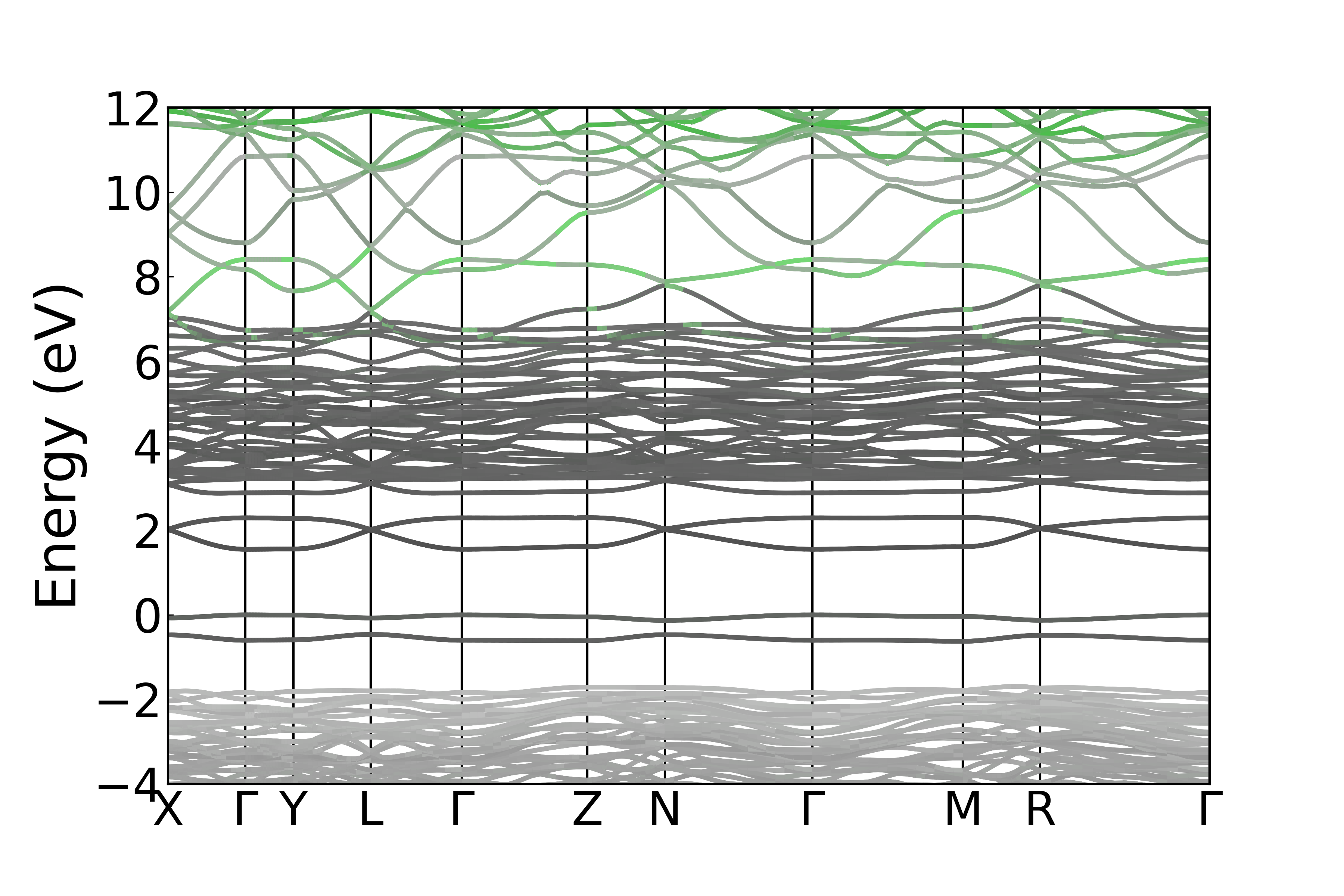}
    \caption{Calculated bandstructure of $\alpha$-Mg$_{0.5}$V$_2$O$_5$ with AFM ordering.}
    \label{Mg05alphaAFM}
\end{figure}
\begin{figure}
    \centering
    \includegraphics[width=0.8\columnwidth]{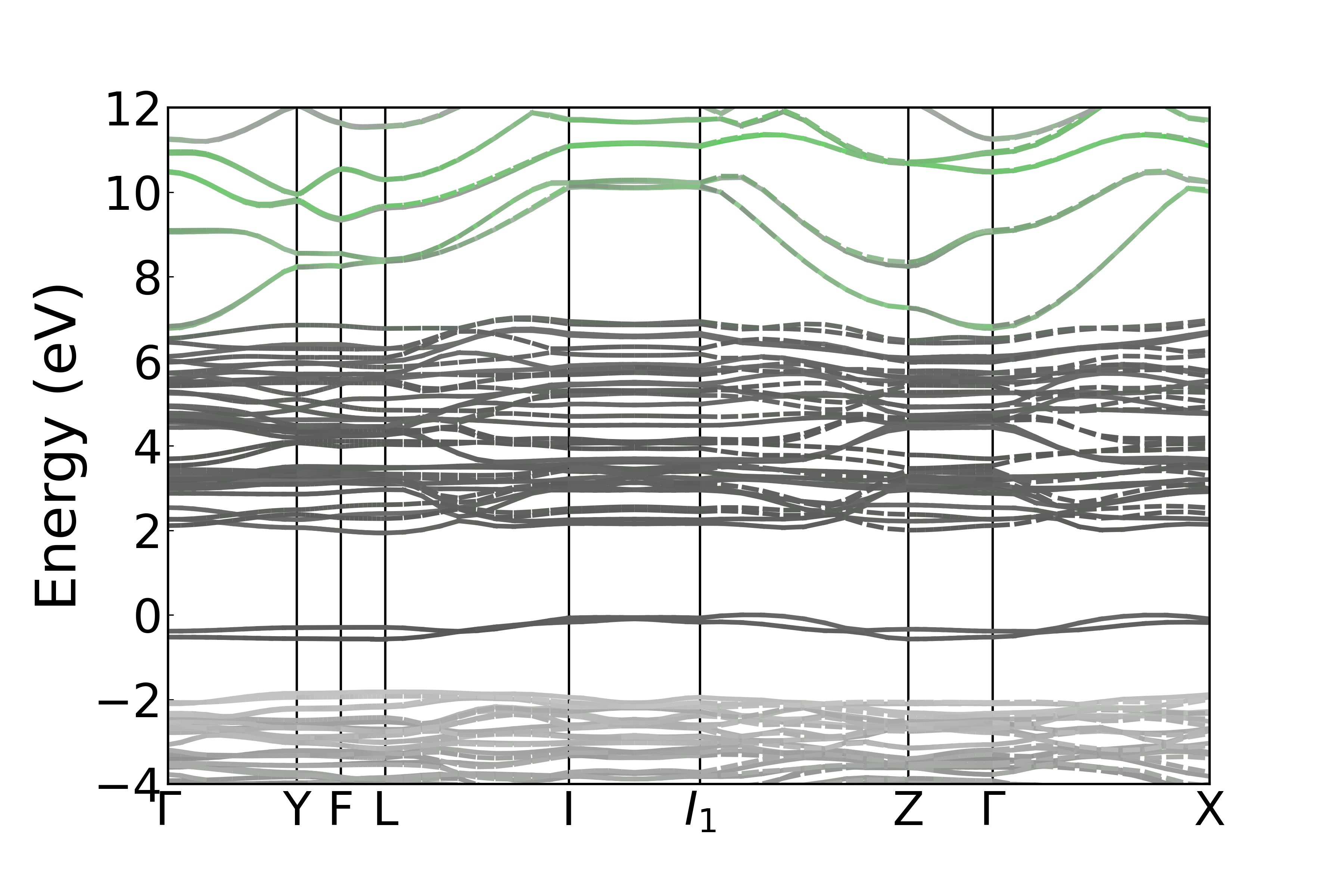}
    \caption{Calculated bandstructure of Mg$_{0.5}$V$_2$O$_5$ in the $\epsilon$-Cu$_{0.85}$ polymorph.}
    \label{Mg05eps}
\end{figure}
\begin{figure}
    \centering
    \includegraphics[width=0.8\columnwidth]{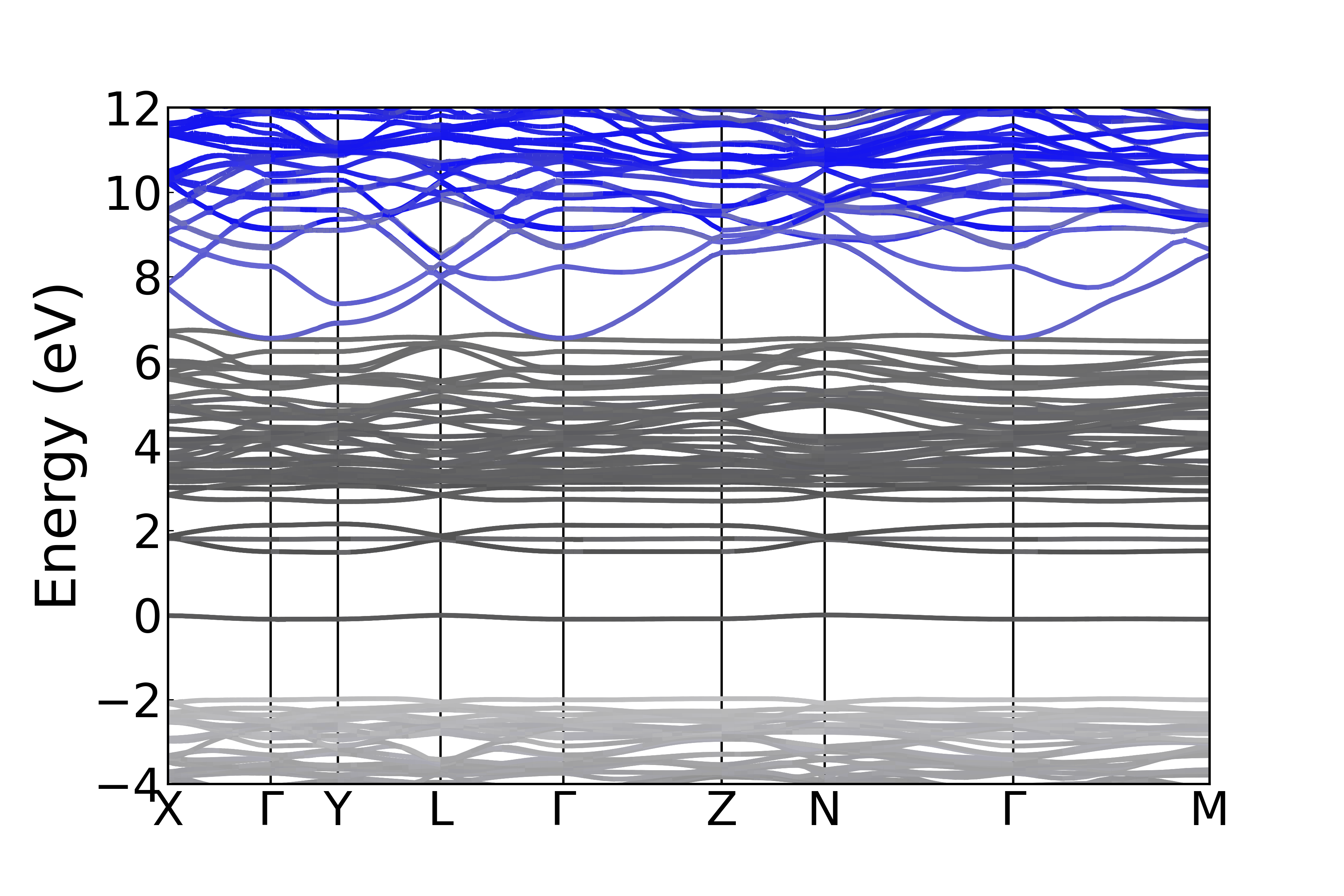}
    \caption{Calculated bandstructure for the $\alpha$-K$_{0.5}$V$_2$O$_5$ polymorph with AFM ordering.}
    \label{K05alphaAFM}
\end{figure}
\begin{figure}
    \centering
    \includegraphics[width=0.8\columnwidth]{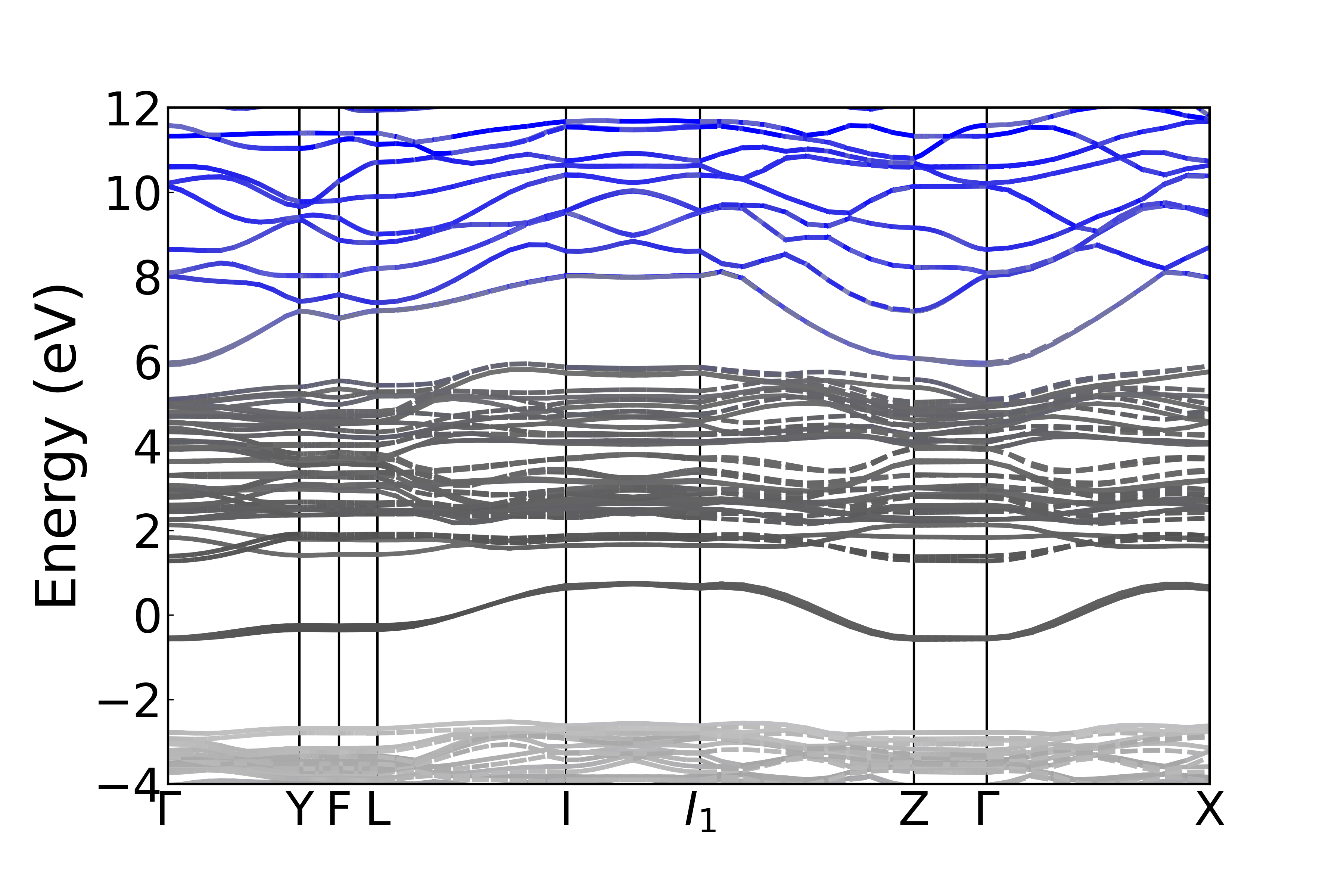}
    \caption{Calculated bandstructure for K$_{0.5}$V$_2$O$_5$ in the $\epsilon$-Cu$_{0.85}$ polymorph.}
    \label{K05eps}
\end{figure}
\begin{figure}
    \centering
    \includegraphics[width=0.8\columnwidth]{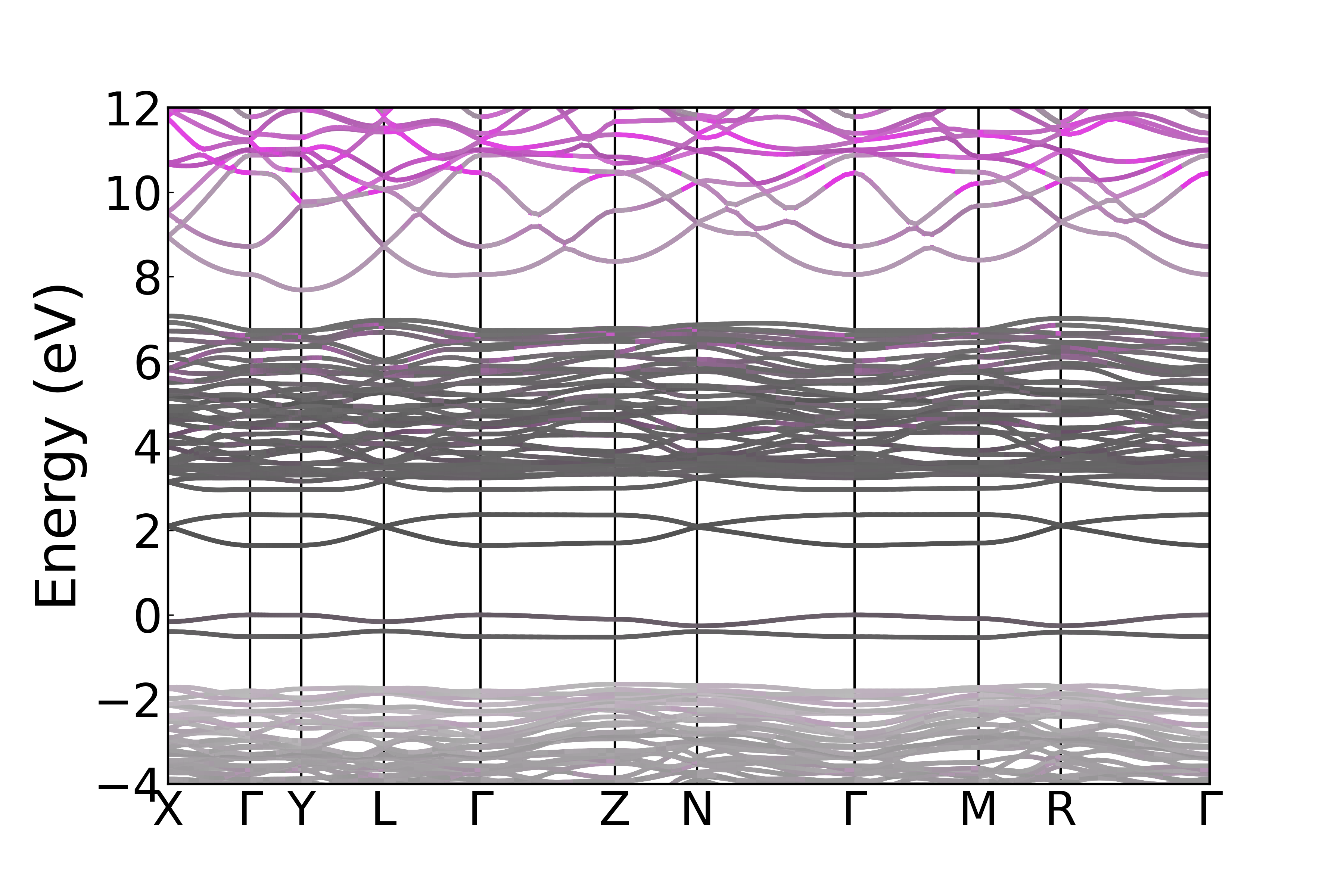}
    \caption{Calculated bandstructure for the $\alpha$-Zn$_{0.5}$V$_2$O$_5$ polymorph with AFM ordering.}
    \label{Zn05alphaAFM}
\end{figure}
\begin{figure}
    \centering
    \includegraphics[width=0.8\columnwidth]{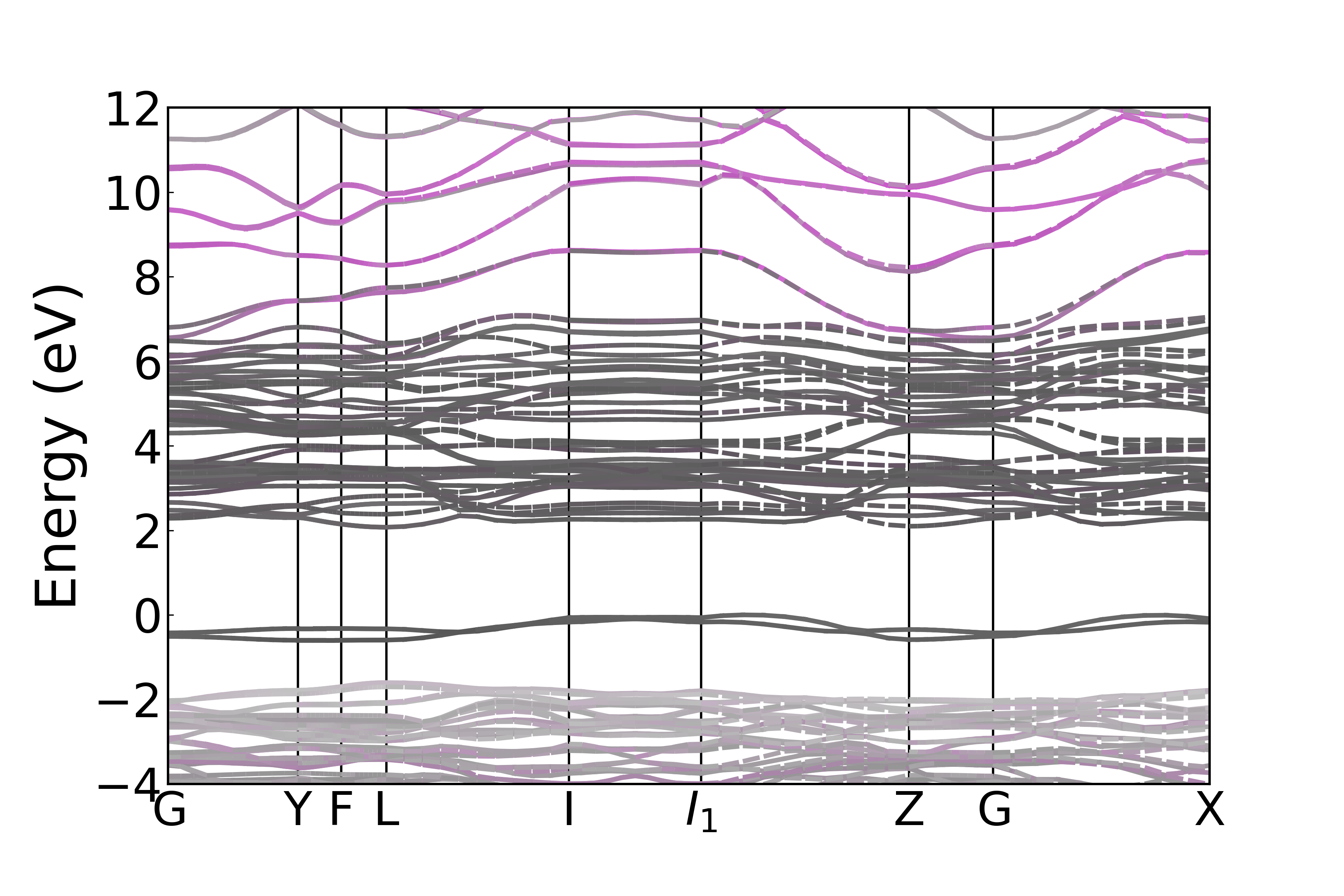}
    \caption{Calculated bandstructure for Zn$_{0.5}$V$_2$O$_5$ in the $\epsilon$-Cu$_{0.85}$ polymorph.}
    \label{Zn05eps}
\end{figure}
\end{document}